\documentclass[12pt,draftclsnofoot,onecolumn]{IEEEtran}

\usepackage{amsmath}
\usepackage{amssymb}
\usepackage{mathtools}

\usepackage{kotex} 
\usepackage[T1]{fontenc}
\usepackage{amsmath, amsfonts, amssymb, amsthm}
\usepackage{multicol}
\usepackage{cite}
\usepackage[pdftex]{graphicx}
\usepackage{color}
\usepackage{cases}
\usepackage{url}
\usepackage{bm}
\usepackage{algorithm}
\usepackage{algpseudocode}

\algnewcommand{\Inputs}[1]{%
  \State \textbf{Inputs:}
  \Statex \hspace*{\algorithmicindent}\parbox[t]{.8\linewidth}{\raggedright #1}
}
\algnewcommand{\Initialize}[1]{%
  \State \textbf{Initialize:}
  \Statex \hspace*{\algorithmicindent}\parbox[t]{.8\linewidth}{\raggedright #1}
}
\algnewcommand{\Iterate}[1]{%
  \State \textbf{Iterate:}
  \Statex \hspace*{\algorithmicindent}\parbox[t]{.8\linewidth}{\raggedright #1}
}
\algnewcommand{\Update}[1]{%
  \State \textbf{Update:}
  \Statex \hspace*{\algorithmicindent}\parbox[t]{.8\linewidth}{\raggedright #1}
}

\allowdisplaybreaks

\newtheorem{theorem}{Theorem}

\newtheorem{lemma}[theorem]{Lemma}
\newtheorem{proposition}{Proposition}

\ifCLASSOPTIONonecolumn
  \newcommand{\figwidth}{0.6\columnwidth}

\else
    
  \newcommand{\figwidth}{.98\columnwidth}

\fi

\begin{document}

\title{Joint Design of Power Control and Access Point Scheduling for Uplink Cell-Free Massive MIMO Networks}

\author{Hyeonsik~Yeom,~\IEEEmembership{Member,~IEEE,}
        Junguk~Park,~\IEEEmembership{Member,~IEEE,}
        Jinho~Choi,~\IEEEmembership{Senior Member,~IEEE,}
        and~Jeongseok~Ha,~\IEEEmembership{Senior Member,~IEEE}

\thanks{ 
    
    H. Yeom, J. Park and J. Ha are with the School of Electrical Engineering, Korea Advanced Institute of Science and Technology (KAIST), Daejeon 34141, South Korea, e-mail:overlimit@kaist.ac.kr, pjuk@kaist.ac.kr, jsha@kaist.edu. 
    
    J. Choi is with the School of Information Technology, Deakin University, Burwood, VIC 3125, Australia, e-mail: jinho.choi@deakin.edu.au. }

 }


\maketitle


\begin{abstract}
This work proposes a joint power control and access points (APs) scheduling algorithm for uplink cell-free massive multiple-input multiple-output (CF-mMIMO) networks without channel hardening assumption. Extensive studies have done on the joint optimization problem assuming the channel hardening. However, it has been reported that the channel hardening may not be validated in some CF-mMIMO environments. In particular, the existing Use-and-then-Forget (UatF) bound based on the channel hardening often seriously underestimates user rates in CF-mMIMO. Therefore, a new performance evaluation technique without resorting to the channel hardening is indispensable for accurate performance estimations. Motivated by this, we propose a new bound on the achievable rate of uplink CF-mMIMO. It is demonstrated that the proposed bound provides a more accurate performance estimate of CF-mMIMO than that of the existing UatF bound. The proposed bound also enables us to develop a joint power control and APs scheduling algorithm targeting at both improving fairness and reducing the resource between APs and a central processing unit (CPU). We conduct extensive performance evaluations and comparisons for systems designed with the proposed and existing algorithms. The comparisons show that a considerable performance improvement is achievable with the proposed algorithm even at reduced resource between APs and CPU.
\end{abstract}



\section{Introduction}
  
\IEEEPARstart{I}{n} the future wireless communication systems, it is an important issue to provide uniformly good quality of service (QoS) throughout the coverage area. Cell-free massive multiple-input multiple-output (CF-mMIMO) system was proposed as a promising network structure to meet the requirement in the next-generation communication \cite{Ngo15Cell-Free,Nayebi15Cell-Free,Ngo17Cell-Free}. In a CF-mMIMO network, a massive number of access points (APs) are distributed across the network, and each AP communicates with a central processing unit (CPU) through a fronthaul link. In addition, all APs in the network cooperatively serve all users that belong to the network with the same communication resources. Thus, no user suffers from the cell-edge effect in a CF-mMIMO network \cite{Demir21Foundations}. It has been widely believed that the features of CF-mMIMO network allow it to enjoy the following advantages over small-cell networks: 1) macro-diversity due to the nature of distributed antennas; 2) no cell-edge effect due to the cooperation of APs; 3) channel hardening effect resulting from the cooperative beamforming with the massive number of APs \cite{Ngo17Cell-Free}.

Meanwhile, the advantages of CF-mMIMO network can be obtained with well an orchestrated resource allocation. In particular, power control is especially important in CF-mMIMO networks, and thus there have been extensive studies on the power control \cite{Ngo17Cell-Free, Nayebi17Precoding, Ngo18OnTheTotal, Nikbakht20Uplink, Guenach21Joint}. Ngo \emph{et al.} in \cite{Ngo17Cell-Free} tackled the power control problem by proposing a scheme maximizing the minimum achievable rate based on the conjugate beamforming (CB). In \cite{Nayebi17Precoding}, Nayebi \emph{et al.} proposed a max-min power control scheme based on both the zero-forcing (ZF) precoding and CB, and heuristic and low complexity power control algorithms were also proposed. Interdonato \emph{et al.} in \cite{Interdonato21Enhanced} proposed enhanced conjugate beamforming (ECB) and developed a max-min power control method based on ECB. In \cite{Ngo18OnTheTotal}, a power control scheme was studied with CB in terms of total energy efficiency instead of achievable rate. That is, the power control scheme in \cite{Ngo18OnTheTotal} was devised to maximize the total energy efficiency. In \cite{Nikbakht20Uplink}, the authors proposed a fractional power control scheme that finds a good trade-off between fairness and average rate.    

In the implementation of CF-mMIMO system, there is an inevitable practical issue associated with fronthaul overhead, i.e., the insatiably growing demands on the fronthaul capacity caused by the traffic between the massive number of APs and CPU. Motivated by this practical issue, there have been studies on the achievable rate of CF-mMIMO system with limited-capacity fronthaul link \cite{Bashar19MaxMin, Bashar21Uplink, Guenach21Joint}. In \cite{Guenach21Joint}, a joint max-min power control and AP scheduling with limited fronthaul connections was investigated. Bashar \emph{et al.,} analyzed achievable rate and energy efficiency of uplink (UL) CF-mMIMO system taking into account uniform quantization for the information from APs to CPU in \cite{Bashar19MaxMin, Bashar21Uplink}, respectively. In addition, they considered a certain fronthaul capacity constraint that the sum of quantization bits delivered to CPU is limited. 

The aforementioned studies on CF-mMIMO system are conducted with the channel hardening assumption which enables one to use the Use-and-then-Forget (UatF) bound and makes the performance analysis greatly simplified. However, recent studies \cite{Polegre20Channel, Chen18Channel} have brought up a question about the validation of channel hardening assumption in CF-mMIMO network where antennas are distributed across a network. Polegre \emph{et. al.,} in \cite{Polegre20Channel} also studied the presence of channel hardening under a general channel environment, i.e., spatially correlated Ricean fading channels. They showed that the channel hardening becomes more distinctive when a line-of-sight (LoS) component is dominant. Meanwhile, for the case of Rayleigh fading channel, it is more favorable for the channel hardening to appear when the channel is uncorrelated. In \cite{Chen18Channel}, the degree of the channel hardening in CF-mMIMO network was analyzed with stochastic geometry according to AP density, the number of antennas per AP, and the path loss coefficient. The authors in \cite{Chen18Channel} showed that the channel hardening assumption in CF-mMIMO network may not be asserted when the path-loss exponent is relatively large. In particular, it is demonstrated that the channel hardening is compromised even with a typical value of the path-loss exponent, e.g., 3.7. Note that each in a network is under different channel environments, e.g., the path-loss coefficients depending on the distance from the user to APs. Thus, some users have the channel hardening while other users do not even if they are in the same network. Later, we will show that when the channel gains between a user and APs have similar statistical properties, the channel harden assumption becomes more distinctive.

  \subsection{Motivation and Our Contributions}
As shown in \cite{Polegre20Channel, Chen18Channel}, users in a network have different degrees of channel hardening depending on their channel environments. Thus, if performance analysis and/or parameter optimizations are conducted based on the UatF bound for all the users, we will end up with inaccurate results and/or suboptimal parameters. In particular, when the rates of the users with weak or no channel hardening effect are performed with the UatF bound, the estimates are considerably underestimated. Accordingly, when we utilize the UatF bound to allocate resources for fairness, excessively more resources will be allocated to the users who do not experience enough channel hardening effect. This motivated us to propose a new bound on achievable rate which does not assume the channel hardening. Based on the proposed bound, we develop a power control and AP scheduling algorithm.

The main contributions of our work are summarized as follows:
  \begin{itemize}
    \item We derive a new lower bound on the achievable rate at a non-coherent receiver for UL CF-mMIMO system without relying on the channel hardening assumption. We prove that the proposed lower bound is tighter than the existing lower bound in \cite[Lemma 4]{Caire18OnTheErgodic}, and extensive performance evaluations show that the proposed lower bound is tighter than the UatF bound. In addition, the proposed bound is in a more suitable form to be evaluated, which makes it especially useful to assess system performance and/or optimize system parameters.
        
    \item Based on the proposed lower bound, we propose a new joint power control and AP scheduling algorithm. Although the joint power control and AP scheduling problem was studied in \cite{Guenach21Joint}, the existing algorithm was developed based on the UatF bound. Moreover, the existing algorithm produces a solution with a fixed number of serving APs per user, which, however, will be shown as suboptimal. This is the first work solving the resource allocation problem, i.e., power and fronthaul bandwidth, for CF-mMIMO system without channel hardening assumption, which enables one to allocate the resource in a more efficient way. In addition, the proposed algorithm is formulated to adaptively allocate a different number APs to a user. It is shown that the improved efficiency of resource allocation and flexibility AP allocation in the proposed algorithm lead to noticeable performance improvement. 
    
    \item Comprehensive performance evaluations are conducted to substantiate our claims: 1) the proposed bound estimates the achievable rates of users in a network regardless of the degrees of channel hardening, 2) more efficient resource allocations are possible with the more accurate estimates of rates, 3) better performances are achievable with the resource allocations obtained with the proposed bound. In addition, the comparisons between the joint power control and AP scheduling with the proposed and UatF bound clearly manifest why the one with the UatF bound fails. The lessons from the comparisons can be utilized to other resource allocation problems. 
  \end{itemize}

  \subsection{Organization}

The remainder of the work is organized as follows: we first describe an UL CF-mMIMO system which includes pilot transmission, channel estimation, and UL data transmission in Section \ref{Sec:SystemModel}. In Section \ref{Sec:LowerBounds}, two widely used lower bounds on achievable rate for non-coherent receiver in CF-mMIMO system are introduced. In addition, we propose a new lower bound on the achievable rate and show that the proposed bound is tighter than the existing bounds. Then, in Section \ref{Sec:JointPowAP}, we present a joint power control and AP scheduling algorithm based on the proposed bound. In Section \ref{Sec:NumResult}, extensive performance evaluations for the UL CF-mMIMO system are carried out with the results of the proposed algorithm. The joint power control and AP scheduling is also conducted with the existing algorithm based on the UatF bound. Then, performance comparisons are carried out for the UL CF-mMIMO system with the results from the proposed and existing algorithms, which clearly shows that significant performance improvements are obtained with the proposed algorithm. In addition, the comparisons give us some useful lessons explaining why the resource allocation with the UatF bound fails. Finally, we conclude this work in Section \ref{Sec:Conclusion}.


\section{System Model} \label{Sec:SystemModel}

  \subsection{General Communication Scenario}

    \begin{figure}
      \centering
      \includegraphics[width=\figwidth]{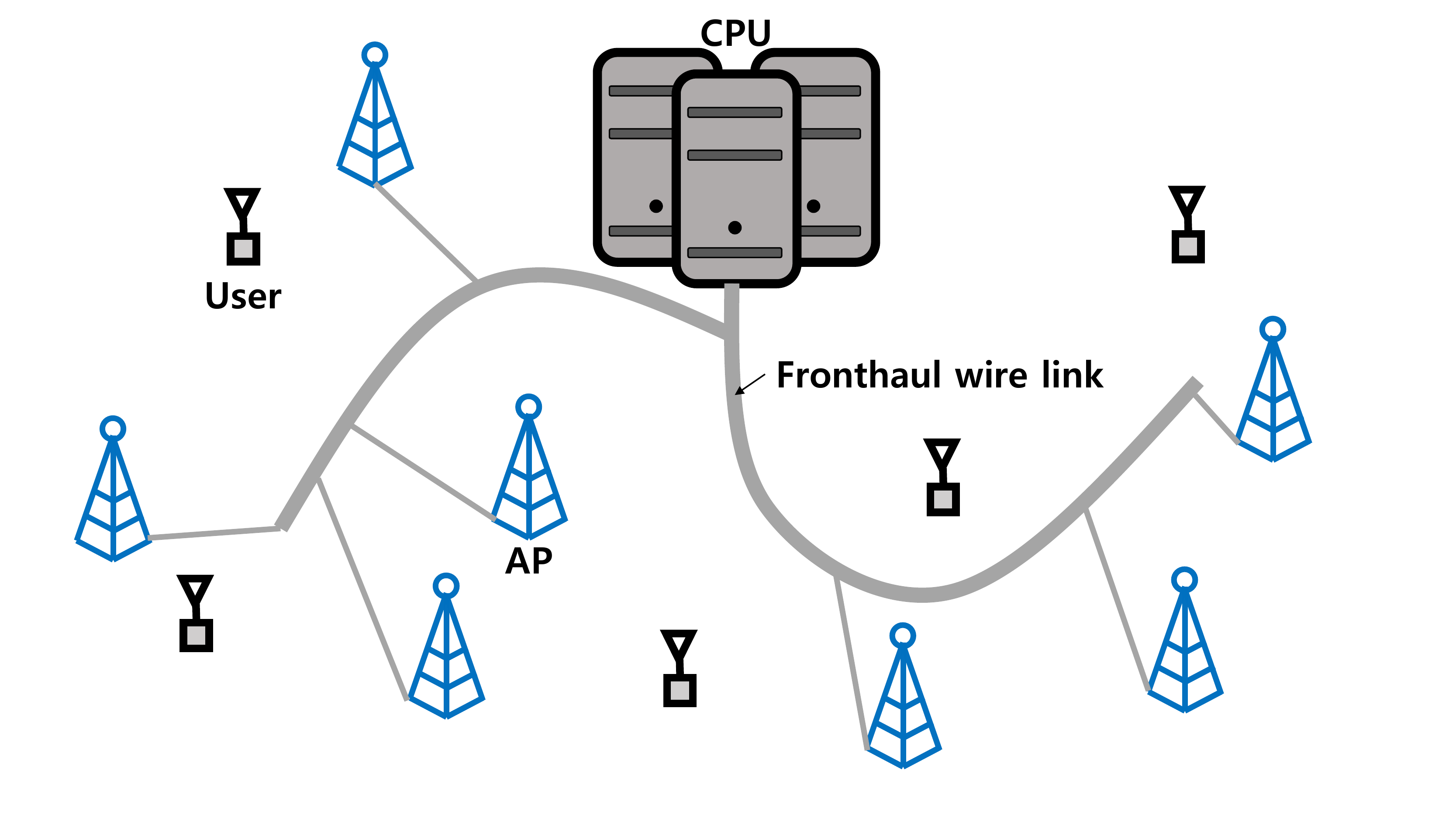}
      \caption{The cell-free massive MIMO network where all APs are connected to CPU via wire fronthaul links.}
      \label{Fig:CF-mMIMO}
    \end{figure}
We consider an UL CF-mMIMO network shown in Fig. \ref{Fig:CF-mMIMO} where there are $M$ APs equipped with $N$ antennas and $K$ single antenna users. Each AP can serve all users by using the same time-frequency resource. Furthermore, all APs are connected via error-free fronthaul links to CPU to deliver received and transmitted user signals. A time-division duplex (TDD) operation with channel reciprocity is assumed. Each coherent duration (in samples) is divided into three parts: UL training, downlink (DL) transmission, and UL transmission \cite{Ngo17Cell-Free}, \cite{Guenach21Joint}. The intervals of the three parts are denoted by $\tau_{\text{p}}, \tau_{\text{d}},$ and $\tau_{\text{u}}$ respectively, and thus, the coherent interval denoted by $\tau_c$ is expressed as the sum of the intervals, i.e., $\tau_{\mathrm{c}} = \tau_{\text{p}} + \tau_{\text{d}} + \tau_{\text{u}}$. Note that due to the channel reciprocity, the UL and DL channels during a coherent time stay the same, which enables one to omit the downlink pilot transmission. In this paper, we focus only on the UL transmission where the fronthaul overhead is significantly larger than that of the DL transmission \cite{Bashar21Uplink}, and thus accurate estimations of user rates are especially crucial for efficient resource allocation. 
    
We consider a block-fading channel which is assumed static over a coherent interval denoted by $\tau_{\mathrm{c}}$. It is also assumed that each codeword is transmitted over a coherent interval and thus experiences the same channel statistics. The channel response between the $m$th AP and the $k$th UE at discrete time $n$ is expressed by $\mathbf{g}_{mk}[n] = \sqrt{\beta_{mk}}  \mathbf{h}_{mk}[n] \in \mathbb{C}^{N \times 1}$ where $\beta_{mk}$ and $\mathbf{h}_{mk}[n] $ indicate the large-scale fading and the small-scale fading, respectively. In this work, we assume that the channel statistics follow the Rayleigh distribution, i.e., $\mathbf{h}_{mk} [n] \sim \mathcal{CN}(\mathbf{0}, \mathbf{I}_{N})$.

  \subsection{Pilot Transmission and Channel Estimation}
In the CF-mMIMO network, each user transmits its own scheduled pilot sequence, and then each AP separately estimates the channel between itself and each user based on the received pilot signal. In this section, the discrete time index, i.e., $n$ is omitted in the description of channel gain since the channel gain is static during the pilot sequences are transmitted. Let $\mathbf{\psi}_i \in \mathbb{C}^{\tau_{\text{p}} \times 1}$ be the $i$th orthonormal pilot sequence. Then, the pilot sequences have the following property.
\[
    \psi_i^T  \psi_j^* =  
        \begin{cases}
            0,  & \text{if} ~ i \neq j, \\
            1,  & \text{Otherwise} ,
        \end{cases}
\]
for all $i, j \in \{1, 2, \ldots , \tau_{\text{p}}\}$. Note that the number of pilot sequences is set to the maximum number of orthogonal sequences in the interval of UL training, i.e., $\tau_{\text{p}}$.

Let us denote by $\mathbf{\varphi}_{k} = \sqrt{\tau_{\text{p}}  \rho_{\text{p}}}  \psi_{i_{k}} \in \mathbb{C}^{\tau_{\text{p}} \times 1}$ the transmitted pilot sequence from the $k$th user, where $\rho_{\text{p}}$ is the transmit power at a user in the training phase, and the symbol, $i_{k} \in \{ 1, 2, \ldots, \tau_{\text{p}} \}$ represents the index of pilot sequence used by the $k$th user. Then the received pilot signal at the $m$th AP in the UL training phase is given by
  \begin{equation}
    \mathbf{Y}_{\text{p},m} = \sum_{k=1}^{K} \mathbf{g}_{mk}  \varphi_{k}^{T} + \mathbf{W}_{\text{p},m} , \label{Eq:RxPilotSig}
  \end{equation}
where $\mathbf{W}_{\text{p},m} \in \mathbb{C}^{N \times \tau_{\text{p}}} $ is the additive-white-Gaussian noise (AWGN) matrix whose elements are independent and identically distributed (i.i.d.)  $\mathcal{CN}(0,\sigma_{\text{p}}^{2})$.
    
The $m$th AP projects the received signal in \eqref{Eq:RxPilotSig} onto $\varphi_k$ to estimate the channel bewteen the $m$th AP and the $k$th user, i.e., $\mathbf{g}_{mk}$ by utilizing a linear minimum mean-square-error (LMMSE) estimator with the knowledge of the channel statistics \cite{Ngo17Cell-Free},\cite{Guenach21Joint},\cite{Bashar21Uplink}. The projected signal at the $m$th AP to estimate the $\mathbf{g}_{mk}$ is expressed as 
\ifCLASSOPTIONonecolumn
\[
    \mathbf{y}_{\text{p},mk} = \mathbf{Y}_{\text{p},m}  \mathbf{\varphi}_{k}^{*} = \sum_{k'=1}^{K} \mathbf{g}_{mk'}  \varphi_{k'}^{T} \varphi_{k}^{*}+ \mathbf{W}_{\text{p},m}  \mathbf{\varphi}_{k}^{*}. \label{Eq:ProjPilotSig}
\]
\else
  \begin{align}
    \mathbf{y}_{\text{p},mk} & = \mathbf{Y}_{\text{p},m}  \mathbf{\varphi}_{k}^{*} \nonumber \\
        & = \sum_{k'=1}^{K} \mathbf{g}_{mk'}  \varphi_{k'}^{T} \varphi_{k}^{*}+ \mathbf{W}_{\text{p},m}  \mathbf{\varphi}_{k}^{*}. \label{Eq:ProjPilotSig}
  \end{align}
\fi
Then, by applying the LMMSE estimation to \eqref{Eq:ProjPilotSig}, we obtain the estimate of $\mathbf{g}_{mk}$ as follows: 
  \begin{align}
    \hat{\mathbf{g}}_{mk} & = \mathbb{E}[\mathbf{g}_{mk}  \mathbf{y}_{\text{p},mk}^{H}]  (\mathbb{E}[\mathbf{y}_{\text{p},mk}  \mathbf{y}_{\text{p},mk}^{H}])^{-1}  \mathbf{y}_{\text{p},mk} \nonumber\\
    & = \frac{\sqrt{\tau_{\text{p}}  \rho_{\text{p}}}  \beta_{mk}}{\tau_{\text{p}}  \rho_{\text{p}}  \beta_{mk} +  \tau_{\text{p}}  \rho_{\text{p}} \sum_{k' \neq k} \beta_{mk'} |\varphi_{k'}^{T} \varphi_{k}^{*}|^2+\sigma_{\text{p}}^{2}} \mathbf{y}_{\text{p},mk}. \label{Eq:EstiChan}
  \end{align}

The estimated channel vector in \eqref{Eq:EstiChan} follows a circularly symmetric complex Gaussian vector with zero mean and covariance matrix $\Sigma_{mk}$, i.e., $\hat{\mathbf{g}}_{mk} \sim \mathcal{CN}(0,\Sigma_{mk})$. The covariance matrix is expressed as \cite{Ngo18OnTheTotal}
\[
\Sigma_{mk} = \mathbb{E} [\hat{\mathbf{g}}_{mk} \hat{\mathbf{g}}_{mk}^{H}] = \gamma_{mk}  \mathbf{I}_{N},
\]
where 
  \begin{equation}
    \gamma_{mk} = \frac{\tau_{\text{p}}  \rho_{\text{p}}  \beta_{mk}^{2}}{\tau_{\text{p}}  \rho_{\text{p}} \beta_{mk} +  \underbrace{\tau_{\text{p}}\rho_{\text{p}} \sum_{k' \neq k} \beta_{mk'} |\varphi_{k'}^{T} \varphi_{k}^{*}|^2}_{\rm (a)} +\sigma_{\text{p}}^{2}}. \label{Eq:VarEstiChan}
  \end{equation}
In \eqref{Eq:VarEstiChan}, the term (a) represents the pilot contamination which is nonzero when the number of users, $K$, is larger than the UL training interval, $\tau_{\rm p}$. However, in this work, we assume that $K \le \tau_{\rm p}$, and thus there is no pilot contamination in \eqref{Eq:VarEstiChan}.

    \subsection{Uplink Data Transmission} \label{Subsec:ULdataTrans}

In the UL data transmission phase, each user transmits its own information signal to all APs with the same time-frequency resource. The transmitted data signal from the $k$th user per the coherent interval is given by 
\[
        x_{k} [n] = \sqrt{\rho_{\text{u}}  \eta_{k}}  q_k [n], \text{ for } 1\le n \le \tau_{\rm u} \nonumber
\]
where $\rho_{\text{u}}$ is the maximum transmit power per user in the UL data transmission, $\eta_{k}$ is the power control coefficient, i.e., $0 \leq \eta_{k} \leq 1, \forall k$, and $q_k$ denotes the information signal transmitted from the $k$th user. The signal, $q_k$ follows the normalized circularly symmetric complex Gaussian distribution with zero mean, i.e., $q_k \sim \mathcal{CN}(0,1)$.
    
Each AP receives a signal, which is the linear combination of the transmitted signals from users and AWGN signal. The received signal at the $m$th AP for a coherent  interval, denoted by  $\mathbf{Y}_m = [\mathbf{y}_m[1], \mathbf{y}_m[2], \ldots, \mathbf{y}_m[\tau_{\text{u}}]] \in \mathbb{C}^{N \times \tau_{\text{u}}}$, is expressed as
\ifCLASSOPTIONonecolumn
  \begin{equation}
    \mathbf{y}_m [n] = \sum_{k=1}^{K} \mathbf{g}_{mk}  x_{k} [n] + \mathbf{w}_{m} [n] = \sum_{k=1}^{K} \sqrt{\rho_{\text{u}}  \eta_{k}}  \mathbf{g}_{mk}  q_{k} [n] + \mathbf{w}_{m} [n], \label{Eq:RxSig}
  \end{equation}
\else
  \begin{align}
    \mathbf{y}_m [n] & = \sum_{k=1}^{K} \mathbf{g}_{mk}  x_{k} [n] + \mathbf{w}_{m} [n] \nonumber \\
    & = \sum_{k=1}^{K} \sqrt{\rho_{\text{u}}  \eta_{k}}  \mathbf{g}_{mk}  q_{k} [n] + \mathbf{w}_{m} [n], \label{Eq:RxSig}
  \end{align}
\fi
where $\mathbf{w}_{m}[n] \sim \mathcal{CN} (0,\sigma_{\text{w}}^{2} \mathbf{I}_N)$ represents the AWGN signal at the $m$th AP. For decoding the $k$th user signal, i.e., $q_k$, a cluster of APs multiply precoding vectors by their received signals, and the resulting signals are delivered to CPU. Then, CPU decodes the $k$th user signal from the signals passed by the cluster of APs. Recently, various schemes to make the precoding vectors have been proposed and analyzed in \cite{Nayebi17Precoding, Interdonato21Enhanced, Bjornson20Scalable} among which CB in \cite{Nayebi17Precoding} has been widely adopted due to its simplicity and reasonable performance. Thus, we also assume CB as the precoding vector in this work. With CB, the received signal for $q_{k}$ at CPU is given by 
\ifCLASSOPTIONonecolumn
  \[
    r_k [n] = \sum_{m=1}^{M} c_{mk}  \mathbf{v}_{mk}^H  \mathbf{y}_m [n] = \sum_{m=1}^{M} c_{mk}  \mathbf{\hat{g}}_{mk}^H  \mathbf{y}_m [n] = \sum_{k'=1}^{K} f_{k,k'}  q_{k'} [n] + n_k [n],  
  \] 
\else
  \begin{align*}
    r_k [n] & = \sum_{m=1}^{M} c_{mk}  \mathbf{v}_{mk}^H  \mathbf{y}_m [n]  \\ 
    & = \sum_{m=1}^{M} c_{mk}  \mathbf{\hat{g}}_{mk}^H  \mathbf{y}_m [n]  \\
    &= \sum_{k'=1}^{K} f_{k,k'}  q_{k'} [n] + n_k [n],  
  \end{align*} 
\fi
where 
  \begin{equation}
     f_{k,k'} \triangleq \sum_{m=1}^{M} \sqrt{\rho_{\text{u}}  \eta_{k'}}  c_{mk}  \mathbf{\hat{g}}_{mk}^H   \mathbf{g}_{mk'}, \label{Eq:Effchen}
  \end{equation}
  \[
        n_k [n] \triangleq \sum_{m=1}^{M} c_{mk}  \mathbf{\hat{g}}_{mk}^H  \mathbf{w}_{m} [n],
  \]
and $c_{mk}$ is an AP connection coefficient telling whether the received signal at the $m$th AP is involved in the decoding of the $k$th user data at CPU or not. That is, $c_{mk} = 1$ when the $m$th AP delivers its received signal to CPU for decoding the $k$th user data. Otherwise,  $c_{mk} = 0$. The achievable rate at CPU for the $k$th user data can be expressed as \cite{Caire18OnTheErgodic}
  \begin{align}
    R_{k} = \frac{1}{\tau_{\text{u}}}  I(q_k [1:\tau_{\text{u}}] ; r_k [1:\tau_{\text{u}}]), \label{Eq:ExactRate}
  \end{align}
where $v[1:\tau_{\text{u}}]$ indicates a vector of length $\tau_{\rm u}$, i.e., $v[1:\tau_{\text{u}}] = [ v[1], v[2], \ldots, v[\tau_{\rm u}] ]$, $I(x ; y)$ represents the mutual information between $x$ and $y$ in bits per channel use. Derivation of a closed-form expression for the achievable rate in \eqref{Eq:ExactRate} is still an open problem. In addition, it is computationally prohibitive to evaluate the achievable rate in \eqref{Eq:ExactRate} for sufficiently large $\tau_{\rm u}$. Thus, most of the existing works for CF-mMIMO system have counted on certain bounding techniques to estimate the achievable rate and/or conduct optimizations of resource allocations. 

\section{Proposed Lower Bound on the Achievable Rate} \label{Sec:LowerBounds}

  \subsection{Existing Lower Bounds}

As mentioned in Section \ref{Subsec:ULdataTrans}, a closed-form of the achievable rate in \eqref{Eq:ExactRate} is unknown. Thus, there are a lot of efforts to obtain a tight bound on the achievable rate in \eqref{Eq:ExactRate}. In this section, we first introduce two lower bounds which have been widely adopted for analyzing and designing CF-mMIMO systems.
    
    \subsubsection{UatF Lower Bound}

The UatF bound denoted by $R_{k}^{\text{UatF}}$ is derived based on the channel hardening and given by
  \begin{equation}
    R_{k} \geq R_{k}^{\text{UatF}} = \log_2(1+\text{SINR}_{k}^{\text{UatF}}), \label{Eq:UatFBound}
  \end{equation}
where $\text{SINR}_{k}^{\text{UatF}}$ is expressed in \eqref{Eq:SINRUatF}. The UatF bound in \eqref{Eq:UatFBound} is in a simple and insightful form which allows one to readily obtain useful lessons and efficiently evaluate system performance. This is why the UatF bound has been so popular in the analysis of co-located Massive MIMO systems \cite[Theorem 4.4]{Bjornson17Massive}.  However, in recent researches \cite{Polegre20Channel, Chen18Channel}, it is shown that the channel hardening may not be sufficiently pronounced at some nodes in a CF-mMIMO network. Thus, performance analysis with the UatF bound becomes inaccurate, and system designs based on the resulting inaccurate performance estimates in turn lead to serious degradation of overall performance.

  \begin{figure*}[t!]
    \begin{align}
      \text{SINR}_{k}^{\text{UatF}} = \frac{(\sqrt{\rho_{\text{u}}  \eta_{k}}  \sum_{m=1}^{M} c_{mk}  N  \gamma_{mk} )^2}{ \sum_{k'=1}^{K} \sum_{m=1}^{M} \rho_{\text{u}}  \eta_{k'}  c_{mk}  N  \gamma_{mk}  \beta_{mk'} + \sum_{m=1}^{M} c_{mk}  \gamma_{mk}  N  \sigma_{w}^{2}}. \label{Eq:SINRUatF}
    \end{align}
  \hrulefill
  \end{figure*}
        
\subsubsection{Caire Lower Bound} \label{SubSubSec:Caire}
To address the weakness of the UatF bound, in \cite{Caire18OnTheErgodic}, Caire proposed two lower bounds on the achievable rate that do not rely on the channel hardening assumption. The bound in \cite[Lemma 3]{Caire18OnTheErgodic} is an insightful form, which makes it often adopted in the analysis of CF-mMIMO system \cite{Chen18Channel}. However, this bound has an issue that it may have negative values in the high SNR regime. The author turns around the problem by proposing another bound in \cite[Lemma 4]{Caire18OnTheErgodic}, which however also has a few technical issues: 1) it is hard to carry out numerical evaluations due to the conditional expectation, and 2) it is not suitable to utilize the bound in optimization problems since it is not in an analytic form. The author in \cite{Chen18Channel} analyzed the DL CF-mMIMO system with the their proposed bound in \cite[Lemma 3]{Caire18OnTheErgodic} and showed that the proposed bounds are tighter than the UatF bound, which leads to a noticeable gap between the performance estimates with the bounds in \cite{Caire18OnTheErgodic} and the UatF bound. 
                
    \subsection{Proposed Lower Bound}

\begin{figure*}[t!]
\ifCLASSOPTIONonecolumn
  \begin{align}
    R_{k}^{\text{Caire}} = & \underbrace{\mathbb{E} \left[ \log_2 \left( 1 + \frac{|f_{k,k}|^2}{\sum_{k' \neq k} \mathbb{E} [|f_{k,k'}|^2 | \mathbf{g}_{mk}, \mathbf{\hat{g}}_{mk}] +  \sum_{m=1}^{M} | c_{mk}  \hat{\mathbf{g}}_{mk}^{H}|^2  \sigma_w^2} \right) \right]}_{(\rm a)} \nonumber\\
      & \hspace{0.15\columnwidth}- \frac{1}{\tau_{\text{u}}} \log_2 \left( 1 + \frac{\tau_{\text{u}}  \mathbb{V}[f_{k,k}]}{\sum_{k' \neq k} \mathbb{E} [|f_{k,k'}|^2] + \mathbb{E} [ \sum_{m=1}^{M} | c_{mk}  \hat{\mathbf{g}}_{mk}^{H}|^2 ]  \sigma_w^2} \right) \nonumber \\
      & \hspace{0.3\columnwidth} + \mathbb{E} \left[ \log_2 \left( \sum _{k' \neq k} |f_{k,k'}|^2 +  \sum_{m=1}^{M} \left| c_{mk}  \hat{\mathbf{g}}_{mk}^{H} \right|^2  \sigma_w^2  \right) \right] \nonumber\\
      & \hspace{0.3\columnwidth} - \log_2 \left( \sum_{k' \neq k} \mathbb{E} [|f_{k,k'}|^2] + \mathbb{E} \left[\sum_{m=1}^{M} \left| c_{mk}  \hat{\mathbf{g}}_{mk}^{H} \right|^2 \right]  \sigma_w^2  \right). \label{Eq:CaireBound}
  \end{align}
\else
  \begin{align}
    R_{k}^{\text{Caire}} = & \underbrace{\mathbb{E} \left[ \log_2 \left( 1 + \frac{|f_{k,k}|^2}{\sum_{k' \neq k} \mathbb{E} [|f_{k,k'}|^2 | \mathbf{g}_{mk}, \mathbf{\hat{g}}_{mk}] +  \sum_{m=1}^{M} | c_{mk}  \hat{\mathbf{g}}_{mk}^{H}|^2  \sigma_w^2} \right) \right]}_{(\rm a)} \nonumber\\
      & - \frac{1}{\tau_{\text{u}}} \log_2 \left( 1 + \frac{\tau_{\text{u}}  \mathbb{V}[f_{k,k}]}{\sum_{k' \neq k} \mathbb{E} [|f_{k,k'}|^2] + \mathbb{E} [ \sum_{m=1}^{M} | c_{mk}  \hat{\mathbf{g}}_{mk}^{H}|^2 ]  \sigma_w^2} \right) \nonumber \\
      & + \mathbb{E} \left[ \log_2 \left( \sum _{k' \neq k} |f_{k,k'}|^2 +  \sum_{m=1}^{M} \left| c_{mk}  \hat{\mathbf{g}}_{mk}^{H} \right|^2  \sigma_w^2  \right) \right] - \log_2 \left( \sum_{k' \neq k} \mathbb{E} [|f_{k,k'}|^2] + \mathbb{E} \left[  \sum_{m=1}^{M} \left| c_{mk}  \hat{\mathbf{g}}_{mk}^{H} \right|^2 \right]  \sigma_w^2  \right). \label{Eq:CaireBound}
  \end{align}
\fi
  \hrulefill
\end{figure*}
    
In this section, we propose a new lower bound on the achievable rate in \eqref{Eq:ExactRate} when the non-coherent receiver does not rely on the channel hardening effect. First off, we reformulate the lower bound \cite[Lemma 4]{Caire18OnTheErgodic} for the UL CF-mMIMO network in Fig. \ref{Fig:CF-mMIMO}, and the resulting bound denoted by $R_{k}^{\text{Caire}}$ is expressed in \eqref{Eq:CaireBound} on the top of the next page. As discussed in Section \ref{SubSubSec:Caire}, the bound, $R_{k}^{\text{Caire}}$ in \eqref{Eq:CaireBound} is not in a suitable form to be evaluated and/or to be utilized for optimization due to the conditional expectation, i.e., 
\[
  \sum_{k \neq k'} \mathbb{E} [|f_{k,k'}|^2 | \mathbf{g}_{mk}, \mathbf{\hat{g}}_{mk}].
\]    
The issues motivate us to propose another bound which is not only more suitable to be evaluated but also tighter than the one in \cite[Lemma 4]{Caire18OnTheErgodic}. The proposed bound is given in Proposition \ref{Prop:Bound}.
  \begin{proposition} \label{Prop:Bound}
    The achievable rate in \eqref{Eq:ExactRate} is lower bounded by $R_k^{\text{Prop}}$ in \eqref{Eq:ProposedBound} which is on the top of the next page.
    \begin{figure*}[t!]
      \begin{align}
        R_k^{\text{Prop}} =  \underbrace{\mathbb{E} \left[ \log_2 \left( \frac{|f_{k,k}|^2 + \sum_{k'\neq k} |f_{k,k'}|^2 + \sum_{m=1}^{M} | c_{mk}  \hat{\mathbf{g}}_{mk}^{H}|^2  \sigma_w^2 }{\sum_{k'\neq k} \mathbb{E} [ |f_{k,k'}|^2 ] + \mathbb{E} [ \sum_{m=1}^{M} | c_{mk}  \hat{\mathbf{g}}_{mk}^{H}|^2 ]  \sigma_w^2} \right) \right]}_{(\rm a)} \nonumber \\
        - \underbrace{\frac{1}{\tau_{\text{u}}} \log_2 \left( 1 + \frac{ \tau_{\text{u}}  \mathbb{V} [f_{k,k}] }{\sum_{k'\neq k} \mathbb{E} [ |f_{k,k'}|^2 ] + \mathbb{E} [ \sum_{m=1}^{M} | c_{mk}  \hat{\mathbf{g}}_{mk}^{H}|^2 ]  \sigma_w^2} \right)}_{(\rm b)}. \label{Eq:ProposedBound}
      \end{align}
      \hrulefill
    \end{figure*}
  \end{proposition}

  \begin{IEEEproof} 
    See Appendix \ref{Sec:ProofofProp1}.
  \end{IEEEproof}

\noindent In the proof, we do not assume the channel hardening and thereby, the bound, $R_{k}^{\text{Prop}}$ in \eqref{Eq:ProposedBound} is applicable regardless of the channel hardening. It is witnessed in \eqref{Eq:ProposedBound} that the bound does not include any conditional expectation. Thus, numerical evaluations of the bound can be readily performed. In addition, the bound is tighter than the existing bound, i.e., the one in \eqref{Eq:CaireBound}, which is confirmed in the proof of Lemma \ref{Lemma:Tight}.

  \begin{lemma} \label{Lemma:Tight}
    The proposed bound in \eqref{Eq:ProposedBound} is tighter than the existing one in \eqref{Eq:CaireBound}, i.e., $R_{k}^{\text{Prop}} \ge   R_{k}^{\text{Caire}}$.
  \end{lemma}
    
  \begin{IEEEproof}
Let us express the term (a) in \eqref{Eq:ProposedBound} as 
  \begin{align}
    & \mathbb{E} \left[ \log_2 \left( \frac{|f_{k,k}|^2 + \sum_{k'\neq k} |f_{k,k'}|^2 + \sum_{m=1}^{M} | c_{mk}  \hat{\mathbf{g}}_{mk}^{H}|^2  \sigma_w^2 }{\sum_{k'\neq k} \mathbb{E} [ |f_{k,k'}|^2 ] + \mathbb{E} [ \sum_{m=1}^{M} | c_{mk}  \hat{\mathbf{g}}_{mk}^{H}|^2 ]  \sigma_w^2} \right) \right] \nonumber \\
    &  = \underbrace{\mathbb{E} \left[ \log_2 \left( 1 + \frac{|f_{k,k}|^2}{\sum_{k' \neq k} |f_{k,k'}|^2 +  \sum_{m=1}^{M} | c_{mk}  \hat{\mathbf{g}}_{mk}^{H}|^2  \sigma_w^2} \right) \right]}_{(\rm a)} \nonumber\\
    &  \ + \mathbb{E} \left[ \log_2 \left( \sum_{k' \neq k} |f_{k,k'}|^2 +  \sum_{m=1}^{M} | c_{mk}  \hat{\mathbf{g}}_{mk}^{H}|^2  \sigma_w^2 \right) \right] \nonumber\\
    &  \ - \log_2 \left( \sum_{k' \neq k} \mathbb{E} [|f_{k,k'}|^2] + \mathbb{E} \left[  \sum_{m=1}^{M} \left| c_{mk}  \hat{\mathbf{g}}_{mk}^{H} \right|^2 \right]  \sigma_w^2  \right). \label{Eq:AppendixB_FirstEq}
  \end{align}
Note that the term (a) in \eqref{Eq:AppendixB_FirstEq} is in the form of $\log(1+\frac{1}{x})$ which is a convex function of $x$. Then, if the Jensen's inequality is applied to (a), the proposed bound becomes the same as the existing bound in \eqref{Eq:CaireBound}. Due to the application of Jensen's inequality, the proposed bound turns out to be larger than or equal to the existing bound. 
  \end{IEEEproof}

While the proposed bound is tighter than the existing bound in \eqref{Eq:CaireBound}, it is noteworthy to tell when the proposed bound becomes closer to the achievable rate in \eqref{Eq:ExactRate}, which is carried out by comparing the proposed bound with an upper bound in \cite[Lemma 1]{Caire18OnTheErgodic}. The upper bound is derived by assuming the instantaneous channel gain is perfectly known to the receiver and given by
\ifCLASSOPTIONonecolumn
  \begin{align}
    & R_k  \leq R_k^{\rm{ub}} \nonumber \\
    & = \mathbb{E} \left[ \log_2 \left( 1 + \frac{ |f_{k,k}|^2 }{ \sum_{k'\neq k}  |f_{k,k'}|^2  +  \sum_{m=1}^{M} | c_{mk}  \hat{\mathbf{g}}_{mk}^{H}|^2  \sigma_w^2 } \right) \right] \nonumber \\
    & = \mathbb{E} \Bigg[ \log_2 \Bigg( \frac{ |f_{k,k}|^2  + \sum_{k'\neq k}  |f_{k,k'}|^2  +  \sum_{m=1}^{M} | c_{mk}  \hat{\mathbf{g}}_{mk}^{H}|^2  \sigma_w^2}{ \sum_{k'\neq k}  |f_{k,k'}|^2  +  \sum_{m=1}^{M} | c_{mk}  \hat{\mathbf{g}}_{mk}^{H}|^2  \sigma_w^2 } \Bigg) \Bigg]. \label{Eq:UpperBound}
  \end{align}
\else  
  \begin{align}
    & R_k  \leq R_k^{\rm{ub}} \nonumber \\
    & = \mathbb{E} \left[ \log_2 \left( 1 + \frac{ |f_{k,k}|^2 }{ \sum_{k'\neq k}  |f_{k,k'}|^2  +  \sum_{m=1}^{M} | c_{mk}  \hat{\mathbf{g}}_{mk}^{H}|^2  \sigma_w^2 } \right) \right] \nonumber \\
    & = \mathbb{E} \Bigg[ \log_2 \Bigg( \nonumber \\
    &~~~~~~~~~~ \frac{ |f_{k,k}|^2  + \sum_{k'\neq k}  |f_{k,k'}|^2  +  \sum_{m=1}^{M} | c_{mk}  \hat{\mathbf{g}}_{mk}^{H}|^2  \sigma_w^2}{ \sum_{k'\neq k}  |f_{k,k'}|^2  +  \sum_{m=1}^{M} | c_{mk}  \hat{\mathbf{g}}_{mk}^{H}|^2  \sigma_w^2 } \Bigg) \Bigg]. \label{Eq:UpperBound}
  \end{align}
\fi
From the comparison between the proposed lower bound in \eqref{Eq:ProposedBound} and the upper bound in \eqref{Eq:UpperBound}, we obtain three conditions that make the proposed bound to be tight:
  \begin{enumerate}
    \item As the interval of the UL data transmission $\tau_{\text{u}}$ gets longer, the proposed bound tends to be larger since the term (b) in \eqref{Eq:ProposedBound} diminishes. Note that the increasing rate of the linear function is faster than that of the logarithm function, and thus the term (b) becomes smaller, which in turn makes the bound larger.
  
    \item As the transmit power grows, the proposed bound in \eqref{Eq:ProposedBound} is closer to the upper bound in \eqref{Eq:UpperBound} when $\tau_{\rm u} \gg 1$, and thus the term (b) is negligible. Note that the term (a) in \eqref{Eq:ProposedBound} and the upper bound in \eqref{Eq:UpperBound} are the same except that the denominator in \eqref{Eq:ProposedBound} is the expectation of the one in \eqref{Eq:UpperBound}. Due to the Jensen's inequality, the growing transmit power makes the logarithm of the denominators with and without the expectation closer to each other. That is, the proposed bound becomes tighter.

    \item As the AP density increases, the proposed bound gets tighter since the summations in the denominator of the upper bound tend to the summations of expectations in the denominators of the proposed bound by the law of large numbers. As discussed in 2), it makes the proposed bound becomes tighter.
  \end{enumerate}

\begin{figure}
      \centering
      \includegraphics[width=\figwidth]{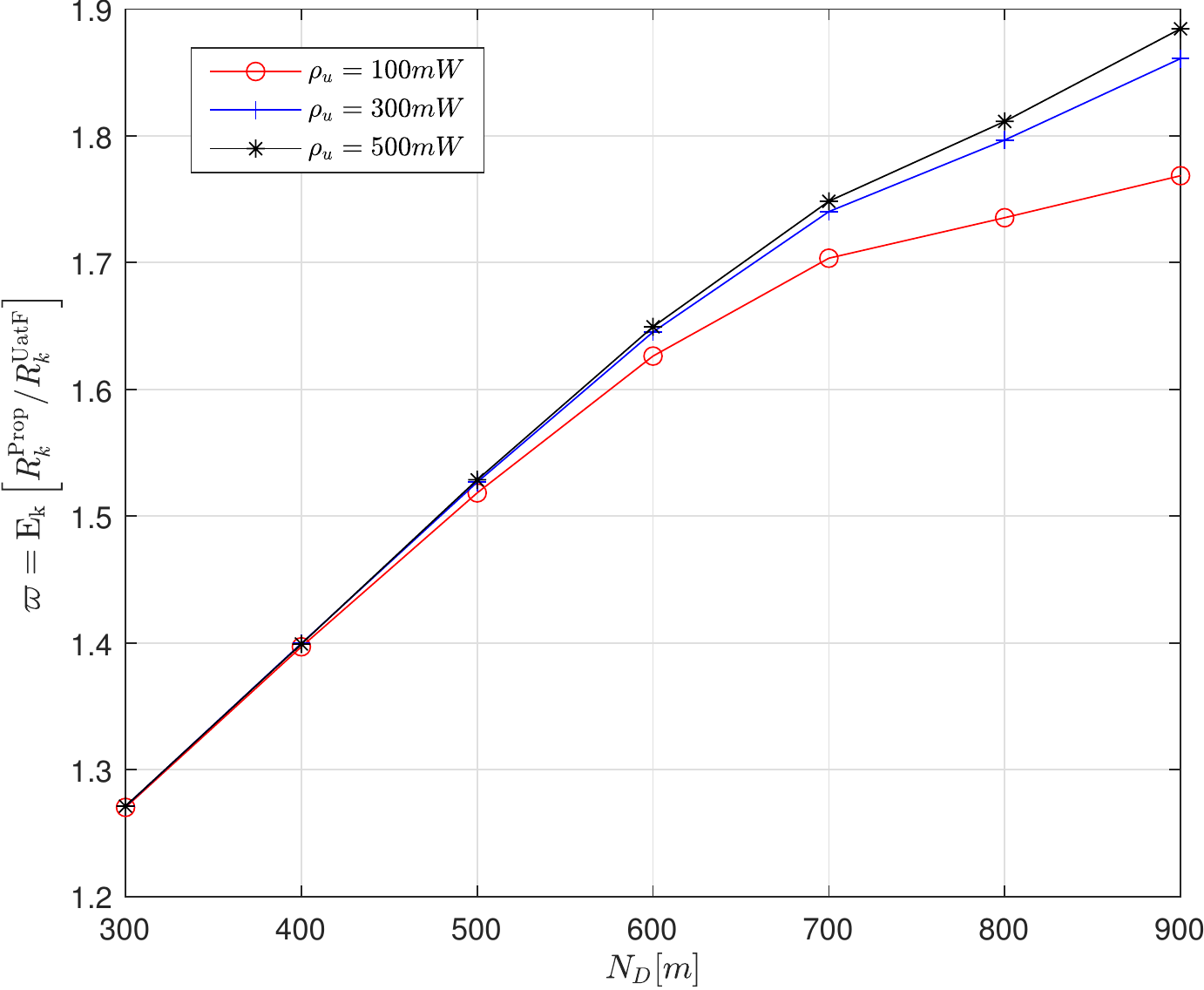}
      \caption{The average ratio between the proposed bound and UatF bound.}
      \label{Fig:Feasible_Ratio}
    \end{figure}

Now, to see the tightness of the proposed bound, we compare the proposed bound with the UatF bound in terms of the average ratio of the proposed bound to the UatF bound, i.e., 
  \begin{equation}
    \varpi = \mathbb{E}_{k} \left[ R_{k}^{\textrm{Prop}}/R_{k}^{\textrm{UatF}} \right] \label{Eq:MeanRarioPropUatF}.
  \end{equation}
Evaluations of the average ratio are performed for various combinations of AP density and transmit power and depicted in Fig. \ref{Fig:Feasible_Ratio}. The  evaluations are conducted assuming the heuristic max-min power control \cite{Nikbakht20Uplink} with a parameter $\vartheta = 1$ and the random network topology in which $K$ users and $M$ APs are randomly distributed across a $D \times D \rm m^2$ square area. The details of other parameters for the evaluation will be introduced in Section \ref{Sec:NumResult}

It is witnessed in Fig. \ref{Fig:Feasible_Ratio} that the measures are larger than unity across all the combinations, which implies that the proposed bound is consistently tighter than the UatF bound. It is noticed in Fig. \ref{Fig:Feasible_Ratio} that as the AP density gets reduced with the increasing $N_D$, the channel hardening becomes weaker, which makes the UatF bound looser and also the average ratio grows. In addition, if the transmit power increases, the average ratio also grows since the proposed bound is closer to the upper bound at high SNR as aforementioned. 

\section{Joint Power Control and AP Scheduling} \label{Sec:JointPowAP}

In this section, we first describe the design problem which maximizes the minimum achievable rate by jointly optimizing the power control and AP scheduling. In particular, we will solve the joint optimization by utilizing the alternating optimization (AO) technique \cite{Guenach21Joint}. 

  \subsection{Problem Forumation} \label{SubSec:OptFrame}
The CF-mMIMO system was originally proposed for guaranteeing uniformly good QoS \cite{Ngo17Cell-Free}. For this reason, most of the existing works considering resource allocation for CF-mMIMO system have taken into account the fairness of performance. To achieve the fairness, the resource allocation problem has been often formulated with the popular max-min measure \cite{Shi14Fairness}. Thus, we also solve the joint power control and AP scheduling problem by formulating an optimization problem with the max-min measure. In doing so, the rates of users are evaluated with the proposed bound \eqref{Eq:ProposedBound} which provides more accurate estimates of the rates regardless of channel hardening and is also computationally amenable. The joint power control and AP scheduling problem is formulated as follows: 
  \begin{subequations}
    \label{Opt:OriginOpt}
    \begin{align}
      \textbf{P1} : \max_{ \{ \eta_{k} \} , \{ c_{mk} \} } & \min_{k} R_k^{\text{Prop}} \\
      \text{s.t.} & \quad 0 \leq \eta_{k} \leq 1, \quad \forall k \label{Eq:PowControl} \\
      & \sum_{m=1}^{M} \sum_{k=1}^{K} c_{mk} \leq \xi_c MK, \label{Eq:FHconst} \\
      & c_{mk} \in \{0,1\}, \quad \forall m,k, \label{Eq:FHBolconst}
    \end{align}
  \end{subequations}
where the parameter, $0 \leq \xi_c \leq 1$, indicates the fronthaul constraint. If the parameter is one, i.e., $\xi_c = 1$, then a user can be served by all APs. 
  
    \begin{figure}
      \centering
      \includegraphics[width=\figwidth]{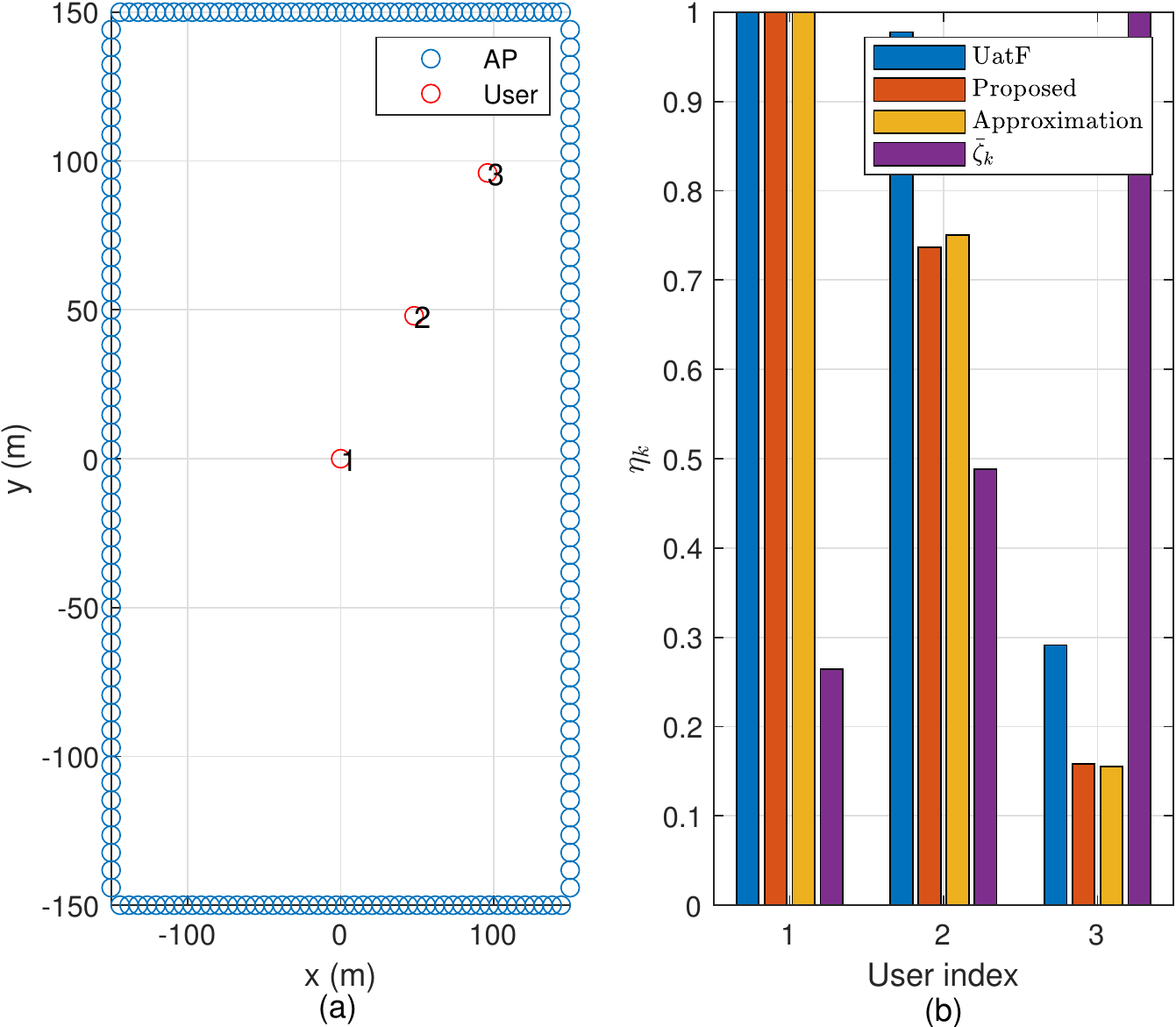}
      \caption{The power control and the normalized channel hardening ratio for $K=3, M=200, N=1$ and $\xi_c = 1$; The results with the UatF, proposed, and approximated bounds are depicted in blue, red and yellow, respectively, and the normalized channel hardening ratio is depicted in purple.}
      \label{Fig:OptPowCont}
    \end{figure}

To confirm the improved accuracy with the proposed bound, also leading to a better design, we consider a toy example with a piazza network shown in Fig. \ref{Fig:OptPowCont}(a) where $K = 3$ single-antenna users are randomly deployed over an area of $D \times D \,\rm m^2 = 300\times300\,\rm m^2$, and $M = 200$ APs equipped with a single antenna ($N = 1$) are located along the boundary of the network area with an equal distance. For the piazza network, we conduct the design in \eqref{Opt:OriginOpt} with the assumption that the sum of $c_{mk}$ in \eqref{Eq:FHconst} holds the equality for $\xi_c = 1$. Thus, the design problem has to be solved only for the power control, which provides the power control coefficients, $\eta_k$'s as its design results. The design is also conducted with the UatF bound, and the design results with both the proposed and UatF bounds are compared in Fig. \ref{Fig:OptPowCont}(b). The ordinate in Fig. \ref{Fig:OptPowCont}(b) indicates either the power control efficient $\eta_k$ or a normalized channel hardening ratio for user $k$, denoted by $\bar \zeta_k$ \cite{Polegre20Channel} which is defined as
\[
  \bar \zeta_{k} = \zeta_k /\max_k[\zeta_k]
\]
where 
\[
  \zeta_{k} = \mathbb{V} \left[ f_{k,k} \right]/\mathbb{E}^2\left[  f_{k,k}  \right] ,
\]
and $f_{k,k}$ is defined in \eqref{Eq:Effchen}. Meanwhile, the abscissa indicates the user index, $k$. Note that the larger the normalized channel hardening ratio is, the weaker the channel hardening becomes. The normalized channel hardening ratio can grow up to unity. The design with the UatF bound is performed with the algorithm introduced in \cite{Ngo17Cell-Free, Guenach21Joint}. Meanwhile, we conduct exhaustive searches for the design with the proposed bound. It is noticed in Fig. \ref{Fig:OptPowCont}(b) that the design results with the UatF bound allocates excessive power to a user of less channel hardening as compared to the one with the proposed bound since the UatF bound underestimates the user rate. The excessive power allocation with the UatF bound leads to considerable degradation of the fairness. However, for the user with the strong channel hardening, i.e., User 1, the power control obtained with the UatF and proposed bounds have the same value.     
  
The results in Fig. \ref{Fig:OptPowCont} clearly show that the proposed bound provides a tighter estimate of user rate and also a better design. However, the proposed bound can be evaluated only in a numerical way and the optimization in \eqref{Opt:OriginOpt} can be solved with exhaustive search, which is not only computationally expensive but fails to provide useful insights. To turn around the technical issue, we propose an approximation of the proposed bound, $R_{k}^{\text{App}}$ which is summarized in Proposition \ref{Prop:ApproxUpp}. 
   
  \begin{proposition} \label{Prop:ApproxUpp}
  The proposed lower bound on the achievable rate in \eqref{Eq:ProposedBound} is approximately upper bounded by $R_{k}^{\text{App}}$ as follows:    
\ifCLASSOPTIONonecolumn
  \begin{equation}
    R_k^{\text{Prop}} \lessapprox R_{k}^{\text{App}} = \log_2 \left( 1 + \text{SINR}_k^{\text{Prop}} \right) - \frac{1}{\tau_{\text{u}}} \log_2 \left( 1 + \frac{\tau_{\text{u}}}{1+MN}  \text{SINR}_k^{\text{Prop}} \right), \label{InEq:ApproxUpper}
  \end{equation}
\else
  \begin{align}
    & R_k^{\text{Prop}} \lessapprox R_{k}^{\text{App}} \nonumber\\
    & = \log_2 \left( 1 + \text{SINR}_k^{\text{Prop}} \right) - \frac{1}{\tau_{\text{u}}} \log_2 \left( 1 + \frac{\tau_{\text{u}}}{1+MN}  \text{SINR}_k^{\text{Prop}} \right), \label{InEq:ApproxUpper}
  \end{align}
\fi
where $\text{SINR}_k^{\text{Prop}}$ is given in \eqref{Eq:SINR_Prop} at the top of the next page. 

\begin{figure*}[t!]
\ifCLASSOPTIONonecolumn
  \begin{align}
    \text{SINR}_k^{\text{Prop}} & = \frac{ \mathbb{E} [|f_{k,k}|^2] }{\sum_{k'\neq k} \mathbb{E} [ |f_{k,k'}|^2 ] + \mathbb{E} [ \sum_{m=1}^{M} | c_{mk}  \hat{\mathbf{g}}_{mk}^{H}|^2 ]  \sigma_w^2} \nonumber \\
    & = \frac{\rho_{\text{u}}  \eta_k  \left[\sum_{m=1}^{M} c_{mk}  N  \gamma_{mk}  \beta_{mk} + \left(\sum_{m=1}^{M} c_{mk}  N  \gamma_{mk} \right)^2 \right] }{\sum_{k' \neq k} \sum_{m=1}^{M} \rho_{\text{u}}  \eta_{k'}  c_{mk}  N  \gamma_{mk}  \beta_{mk'} + \sum_{m=1}^{M} c_{mk}  \gamma_{mk}  N  \sigma_w^2}.
      \label{Eq:SINR_Prop}
  \end{align} \nonumber
\else
  \begin{align}
    \text{SINR}_k^{\text{Prop}} & = \frac{ \mathbb{E} [|f_{k,k}|^2] }{\sum_{k'\neq k} \mathbb{E} [ |f_{k,k'}|^2 ] + \mathbb{E} [ \sum_{m=1}^{M} | c_{mk}  \hat{\mathbf{g}}_{mk}^{H}|^2 ]  \sigma_w^2} 
    = \frac{\rho_{\text{u}}  \eta_k  \left[\sum_{m=1}^{M} c_{mk}  N  \gamma_{mk}  \beta_{mk} + \left(\sum_{m=1}^{M} c_{mk}  N  \gamma_{mk} \right)^2 \right] }{\sum_{k' \neq k} \sum_{m=1}^{M} \rho_{\text{u}}  \eta_{k'}  c_{mk}  N  \gamma_{mk}  \beta_{mk'} + \sum_{m=1}^{M} c_{mk}  \gamma_{mk}  N  \sigma_w^2}.
    \label{Eq:SINR_Prop}
  \end{align} \nonumber
\fi
    \hrulefill
  \end{figure*}
        
  \begin{IEEEproof}
  Refer to Appendix \ref{Sec:ProofofProp2}.
  \end{IEEEproof}    
    \end{proposition}         
The solutions of the design problem in \eqref{Opt:OriginOpt} with the proposed bound and the approximation in \eqref{Prop:ApproxUpp} are compared in Fig.  \ref{Fig:OptPowCont}(b) where both design results look close to each other.  From the approximation in \eqref{InEq:ApproxUpper}, we can obtain a helpful property in Lemma \ref{Lemma:Prop} which simplifies the optimization problem in \eqref{Opt:OriginOpt}.

  \begin{lemma} \label{Lemma:Prop}
    The approximated bound \eqref{InEq:ApproxUpper} is a monotonically increasing function of $\text{SINR}_k^{\text{Prop}}$ in \eqref{Eq:SINR_Prop}. That is,
    \[
    R_{k}^{\text{App}} \propto \text{SINR}_k^{\text{Prop}}. \label{ProptionalTerm}
    \]
  \end{lemma}
  \begin{IEEEproof}
  For simplicity, let $x = \text{SINR}_k^{\text{Prop}}$ and $f(x) = R_{k}^{\text{App}}$. Then, we can express the derivative of $f(x)$ with respect to $x$ as
\ifCLASSOPTIONonecolumn
  \begin{equation}
    \frac{\partial f(x)}{\partial x} = \frac{1}{\ln 2} \left( \frac{1}{x+1} - \frac{1}{x \tau_{\text{u}} + 1 + MN} \right) = \frac{1}{\ln 2} \left( \frac{x  (\tau_{\text{u}} - 1) + MN}{(x+1)(x  \tau_{\text{u}} + 1 + MN)} \right) > 0, \nonumber
  \end{equation}
\else
  \begin{align}
    \frac{\partial f(x)}{\partial x} & = \frac{1}{\ln 2} \left( \frac{1}{x+1} - \frac{1}{x \tau_{\text{u}} + 1 + MN} \right) \nonumber \\
    & = \frac{1}{\ln 2} \left( \frac{x  (\tau_{\text{u}} - 1) + MN}{(x+1)(x  \tau_{\text{u}} + 1 + MN)} \right) > 0, \nonumber
  \end{align}
\fi
  where the last inequality holds since $\tau_{\text{u}} > 1$, $M > 0$, and $N > 0$. Due to the inequality, we can conclude that the approximated bound is a monotonically increasing function of $\text{SINR}_k^{\text{Prop}}$.
  \end{IEEEproof}
  
By utilizing Lemma \ref{Lemma:Prop}, we can reformulate the optimization problem \eqref{Opt:OriginOpt} as
  \begin{subequations}
    \label{Opt:ApproxUp_Opt}
    \begin{align}
      \textbf{P2}: & \max_{ \{ \eta_{k} \}, \{ c_{mk}  \} } \min_{k} \text{SINR}_k^{\text{Prop}}, \\
      & \text{s.t. } \eqref{Eq:PowControl}, \eqref{Eq:FHconst}, \text{ and } \eqref{Eq:FHBolconst}.
    \end{align}
  \end{subequations}
While the objective function is now turned into an analytic form, the optimization problem in \eqref{Opt:ApproxUp_Opt} cannot be efficiently solved. To resolve this problem, we utilize the AO technique in which the optimization is broken into two independent sub-optimization problems, i.e., the power control and AP scheduling problems in this work, and the two optimization problems are solved in the alternating manner. That is, the power control problem is solved with the optimization results of the AP scheduling problem, vice versa. The two optimization problems, i.e.,  the power control and AP scheduling problems are discussed in the following subsections.

  \subsection{Power Control} \label{SubSec:PowCont}

In this section, we describe the optimization problem which finds the UL power control coefficient when the AP scheduling is fixed. Then, we can express the optimization problem as 
  \begin{subequations}
  \label{Opt:PowAll}
    \begin{align}
      \textbf{P3}: \max_{ \{ \eta_{k} \} } & \min_{k} \text{SINR}_k^{\text{Prop}}, \\
      \text{s.t. } &  \eqref{Eq:PowControl}.
    \end{align}
  \end{subequations}
Unfortunately, the optimization problem in \eqref{Opt:PowAll} is not concave but it will be shown quasi-concave. To solve the non-convex optimization problem in \eqref{Opt:PowAll}, we first prove the optimization problem is a quasi-concave optimization problem in the proof of Proposition \ref{Prop:QuasiConcave}.

  \begin{proposition} \label{Prop:QuasiConcave}
    The optimization problem in \eqref{Opt:PowAll} is quasi-concave.
  \end{proposition}

  \begin{IEEEproof}
    See Appendix \ref{Sec:ProofofProp3}.
  \end{IEEEproof}

It is known \cite[Section 4.2.5]{Boyd04Convex}, \cite{Ngo17Cell-Free} that a quasi-concave optimization problem can be solved with a combination of convex optimization and bisection search in which a convex feasibility test is performed at each step.

  \subsection{AP Scheduling} \label{Sec:AP Scheduling}

Like the power control problem in Section \ref{SubSec:PowCont}, we formulate the AP scheduling problem for given power control coefficients as follows:

  \begin{subequations}\label{Opt:APSch_original}
    \begin{align}
      \textbf{P5} : \max_{\substack{ \{c_{mk}\} } } & \min_{\substack{k}} \text{SINR}_k^{\text{Prop}}  \\
      \text{s.t. } & \eqref{Eq:FHconst}, \text{ and } \eqref{Eq:FHBolconst}.
    \end{align}
  \end{subequations}
Unfortunately, the problem in \eqref{Opt:APSch_original} is not quasi-concave. Moreover, it contains the boolean constraint \eqref{Eq:FHBolconst}, which makes it harder to solve the problem. To turn around the issues, we introduce a relaxation, so-called \emph{linear program relaxation} \cite[Section 4]{Boyd04Convex} that replaces the boolean constraint with a continuous constraint. In addition, to resolve the non-concavity, we use the majorization-minimization (MM) algorithm \cite{Sun17Majorization-Minimization} which provides a solution of an optimization problem by iteratively optimizing a surrogate function instead of the objective function.

We first relax the boolean constraint as a continuous value constraint. Then, the optimization problem in \eqref{Opt:APSch_original} turns out to be 
  \begin{subequations}\label{Opt:APSch_Relax}
    \begin{align}
      \textbf{P5} : \max_{\substack{ \{c_{mk}\} } } & \min_{\substack{k}} \text{SINR}_k^{\text{Prop}}  \\
      \text{s.t. } & \eqref{Eq:FHconst}, \\
      &  \sum_{m=1}^{M}  c_{mk} \geq 1, \quad \forall k \label{Eq:AddrelaxConst}\\
      & 0 \leq c_{mk} \leq 1, \quad \forall m,k , \label{Eq:RelaxConst}
    \end{align}
  \end{subequations}
where the constraint in \eqref{Eq:AddrelaxConst} is introduced to prevent the case that there are some users served by less than one AP, and the constraint \eqref{Eq:RelaxConst} is due to the relaxation. Next, we solve the relaxed problem in \eqref{Opt:APSch_Relax} by applying the MM algorithm. To do so, it is necessary to find a surrogate function, $g( \mathbf{c}_{k} ; \mathbf{c}_{k}^{[n]} )$ satisfying 
  \begin{equation} \label{Eq:surrogate}
    g( \mathbf{c}_{k} ; \mathbf{c}_{k}^{[n]} ) \le \text{SINR}_k^{\text{Prop}}(\mathbf c_k )  + \omega_k, \forall\, \mathbf{c}_k
  \end{equation}
where $n$ is the iteration index, $\mathbf{c}_k = [c_{1k}, c_{2k}, \ldots , c_{Mk}]^T \in \mathbb{C}^{M \times 1}$, and  $\omega_n = g( \mathbf{c}_{k} ; \mathbf{c}_{k}^{[n]} ) - \text{SINR}_k^{\text{Prop}}(\mathbf c_k )$. We can find the surrogate function for the objective function $\text{SINR}_k^{\text{Prop}}(\mathbf c_k )$ as follows:
\ifCLASSOPTIONonecolumn
  \begin{equation}
    g( \mathbf{c}_{k} ; \mathbf{c}_{k}^{[n]} ) = \frac{\eta_{k}  ( \bm{\gamma}_{k} \otimes {\boldsymbol{\beta}}_{k})^{T}    \mathbf{c}_{k} + N  \eta_{k}  \left( \left( \bm{\gamma}_{k}^{T}    \mathbf{c}_{k}^{[n]} \right)^2 + 2  \mathbf{c}_{k}^{[n]}  (\mathbf{c}_{k}-\mathbf{c}_{k}^{[n]})\right)}{\sum_{k' \neq k} \eta_{k'}  ( \bm{\gamma}_{k} \otimes {\boldsymbol{\beta}}_{k'})^T  + \frac{\sigma_w^2}{\rho_{\text{u}}} {\boldsymbol{\gamma}}_{k}}^T, \label{Eq:Surro_g}
  \end{equation}
\else
  \begin{align}
    & g( \mathbf{c}_{k} ; \mathbf{c}_{k}^{[n]} ) \nonumber \\ 
    & = \frac{\eta_{k}  ( \bm{\gamma}_{k} \otimes {\boldsymbol{\beta}}_{k})^{T}    \mathbf{c}_{k} + N  \eta_{k}  \left( \left( \bm{\gamma}_{k}^{T}    \mathbf{c}_{k}^{[n]} \right)^2 + 2  \mathbf{c}_{k}^{[n]}  (\mathbf{c}_{k}-\mathbf{c}_{k}^{[n]})\right)}{\sum_{k' \neq k} \eta_{k'}  ( \bm{\gamma}_{k} \otimes {\boldsymbol{\beta}}_{k'})^T  + \frac{\sigma_w^2}{\rho_{\text{u}}} {\boldsymbol{\gamma}}_{k}}^T, \label{Eq:Surro_g}
  \end{align}
\fi
where 
  \begin{align*}
    {\bm{\gamma}}_k & = [\gamma_{1k}, \gamma_{2k}, \ldots , \gamma_{Mk}]^T \in \mathbb{C}^{M \times 1},  \\
\bm{\beta}_k & = [\beta_{1k}, \beta_{2k}, \ldots , \beta_{Mk}]^T \in \mathbb{C}^{M \times 1}.
  \end{align*}
Note that $g( \mathbf{c}_{k} ; \mathbf{c}_{k}^{[n]} )$ in \eqref{Eq:Surro_g} is $\text{SINR}_k^{\text{Prop}}$ except that the numerator of $\text{SINR}_k^{\text{Prop}}$ is replaced with its first order Taylor expansion. Since the numerator is a quadratic function, the first-order Taylor expansion is always less than or equal, which makes the condition in \eqref{Eq:surrogate} satisfied.

The surrogate function in \eqref{Eq:Surro_g} allows us to reformulate the optimization problem in \eqref{Opt:APSch_Relax} at the $n$-th iteration of the MM algorithm as follows:
  \begin{subequations}\label{Opt:APSch_RelaxMM}
    \begin{align}
      \textbf{P5} : \max_{\substack{ \{ \mathbf{c}_{k} \} } } & \min_{\substack{k}} g( \mathbf{c}_{k} ; \mathbf{c}_{k}^{[n]} )  \\
      \text{s.t. } & \eqref{Eq:FHconst}, \eqref{Eq:AddrelaxConst}, \text{ and } \eqref{Eq:RelaxConst}.
    \end{align}
  \end{subequations}
Fortunately, this optimization problem in \eqref{Opt:APSch_RelaxMM} is now quasi-concave, which is explained in the proof of Proposition \ref{Prop:QuasiConcave_APsch}.

  \begin{proposition} \label{Prop:QuasiConcave_APsch}
    The optimization problem in \eqref{Opt:APSch_RelaxMM} is a quasi-concave optimization problem.
  \end{proposition}
  \begin{IEEEproof}
    See Appendix \ref{Sec:ProofofProp4}.
  \end{IEEEproof}
Since the AP scheduling problem is now quasi-concave, it can be readily solved with a combination of convex optimization and bisection search \cite[Section 4.2.5]{Boyd04Convex}, \cite{Ngo17Cell-Free} as discussed in Section \ref{SubSec:PowCont}. It should be noted that the convergence speed of the MM algorithm depends on its initial value, i.e., $ \{ \mathbf{c}_{k}^{[0]} \}$. In this work, we take the initial AP connection coefficients, $c_{mk}$ from a heuristic AP scheduling scheme, i.e., the LLSF (Largest Large-Scale Fading) algorithm in \cite{Guenach21Joint} as the initial value of the MM algorithm.

The resulting AP connection coefficients from the optimization problem in \eqref{Opt:APSch_RelaxMM}, denoted by $\mathbf c^\ast _k = [c^\ast_{1k}, c^\ast_{2k}, \ldots , c^\ast_{Mk}]^T$ for $1 \le k \le K$, are continuous values between zero and one. To convert them into binary values, the rounding technique \cite{Guenach21Joint} is often individually applied to each AP connection coefficient. However, such rounding technique may induce an undesirable event in that the sum of AP coefficients is larger than the total fronthaul constraint, i.e., $\xi_c MK$. To prevent such event, we instead apply the rounding technique in a judicious way that we first round the sum of the continuous values, i.e., $w_k = \lceil \sum_{m=1}^{M} c^\ast_{m k} \rfloor$, and set $c^\ast_{m_i k}$ to one for $1 \le i \le w_k$ such that $c^\ast_{m_i k} \ge c^\ast_{m_j k}$ for $i < j$. The algorithm converting the continuous into discrete AP connection coefficients is summarized in Algorithm \ref{Alg:Converting Continuous into Discrete AP Coefficient}.

  \begin{algorithm}
  \caption{Converting Continuous into Discrete AP Connection Coefficients}
  \label{Alg:Converting Continuous into Discrete AP Coefficient}
    \begin{algorithmic}[1]
      \For{$k = 1$ do $K$}
        \State{$w_{k} = \lceil \sum_{m=1}^{M}c_{m k} \rfloor$}
        \State{Sorting the $\mathbf{c}_{k}^{*}$ in the descending order:  $$c_{m_1 k}^{*} \geq c_{m_2 k}^{*} \cdots \geq c_{m_M k}^{*}$$}
        \For{$m = 1$ do $M$}
          \If{ $m \in \{ m_1, m_2, \ldots, m_{w_k} \}$ }
            \State{Set $c_{m k} = 1$}
          \Else
            \State{Set $c_{m k} = 0$}
          \EndIf
        \EndFor
     \EndFor      
    \end{algorithmic}
  \end{algorithm}

  \subsection{Joint Allocation with the Alternative Optimization}

The purpose of this work is to design a CF-mMIMO system with joint optimization of power control and AP scheduling, which however is numerically too much demanding. To turn around the technical obstacle, we adopt the AO technique \cite{Zhou22Fairness, Guenach21Joint, Zhao22Joint} with the power control and AP scheduling as two independent sub-optimization problems. The alternation between the two optimizations is repeated until the objective function is saturated. The joint optimization for power control and AP scheduling is summarized in Algorithm \ref{Alg:JointAOFrame}.

  \begin{algorithm}
  \caption{Joint Power Control and AP Scheduling with the Alternative Optimization}
  \label{Alg:JointAOFrame}
    \begin{algorithmic}[1]
      \State Initialize $n \gets 0$, $ \{ \mathbf{c}_{k}^{[0]} \} $, $\epsilon$
      \Repeat
        \State $n \gets n+1$
        \State Given $ \{ \mathbf{c}_{k}^{[n-1]} \}$, solve \eqref{Opt:PowAll}, and get $ \{ \eta_{k}^{*} \} $
	\State Given $\{ \eta_{k}^{*} \}$, solve \eqref{Opt:APSch_Relax}, and get $ \{ \mathbf{c}_{k}^{*} \} $
        \State Update $ \{ \eta_{k}^{[n]} \} \leftarrow \{ \eta_{k}^{*} \} $ and $ \{ \mathbf{c}_{k}^{[n]} \} \leftarrow \{ \mathbf{c}_{k}^{*} \} $
      \Until { With $h(\bm{\eta}, \mathbf{C}) = \min_{\substack{k}} \text{SINR}_k^{\text{Prop}} (\bm{\eta}, \mathbf{C})$,
$$ \left|  h (\bm{\eta}^{[n+1]},\mathbf{C}^{[n+1]}) - h (\bm{\eta}^{[n]},\mathbf{C}^{[n]}) \right| \leq \epsilon$$  }
      \State Using Algorithm \ref{Alg:Converting Continuous into Discrete AP Coefficient}, convert the continuous values of $\mathbf{c}_{k}^{[n]}$ into discrete ones.
    \end{algorithmic}
  \end{algorithm}
Since the convergence of the alternative optimization is vital, we will discuss the convergence of the joint optimization for power control and AP scheduling using the AO technique. First off, let $h(\bm{\eta}, \mathbf{C}) = \min_{\substack{k}} \text{SINR}_k^{\text{Prop}} (\bm{\eta}, \mathbf{C})$ where $\bm{\eta} = [ \eta_1, \eta_2, \ldots, \eta_K ]^T$ and $\mathbf{C} = [\mathbf{c}_{1}, \mathbf{c}_{2}, \ldots, \mathbf{c}_{K}]$. Then, it can be readily confirmed that the following inequalities hold:
\[
	h(\bm{\eta}^{[n]},\mathbf{C}^{[n]}) \le h(\bm{\eta}^{[n+1]},\mathbf{C}^{[n]}) \le h(\bm{\eta}^{[n+1]},\mathbf{C}^{[n+1]}), \forall n \in \mathbb{Z}  
\]
where $\bm{\eta}^{[n]}$ and $\mathbf{C}^{[n]}$ indicate the results of the proposed power control and AP scheduling algorithms discussed in Sections \ref{SubSec:PowCont}, and \ref{Sec:AP Scheduling}, respectively, at the $n$th alternation in the joint optimization. If we define a sequence, $ \{ a_n  \} $ where $a_n = h(\bm{\eta}^{[n]},\mathbf{C}^{[n]})$, then the sequence $ \{ a_n  \} $ is monotone increasing. Moreover, the sequence is bounded since the communication resources, i.e., the transmit power and the fronthaul bandwidth, are constrained. Therefore, by using the monotone convergence theorem \cite{Copson70OnaGeneralisation}, we can conclude that the objective function converges as the joint optimization proceeds.

\section{Numerical Results} \label{Sec:NumResult}

In this section, we evaluate performances of CF-mMIMO network in Fig. \ref{Fig:CF-mMIMO} in terms of throughput and fronthaul usage based on the proposed and the existing UatF bounds. The fronthaul usage is defined as follows:
\begin{equation} \label{eq:usage}
  \Xi = \sum_{m=1}^{M} \sum_{k=1}^{K} c_{mk}  
\end{equation}
which tells the sum of serving APs for all users. The performance evaluations are carried out with both the proposed and existing \cite{Guenach21Joint} power control and/or AP scheduling when the number of user served by an AP is limited. Finally, we demonstrate the convergence of the proposed Algorithm \ref{Alg:JointAOFrame}.

  \subsection{Simulation Setup}

For the performance evaluations, we consider two cases of practical interest, i.e., the random \cite{Ngo17Cell-Free} and piazza networks \cite{Guenach21Joint, Interdonato19Ubiquitous}. In both cases, $K$ single-antenna users are randomly deployed with uniform distribution in a square network with $D \times D \mathrm{m^2}$ area.  In the random network, $M$ APs equipped with $N$ antenna are randomly deployed following the uniform distribution. That is, all nodes in the network are distributed by the binomial point process (BPP).  Moreover, we also use the wraparound technique \cite{Ngo17Cell-Free} to prevent the edge effect in the random network. Meanwhile, in the piazza network, all APs are located in the boundary of the network with an equal distance. The performance evaluations are carried out with $M=200$ and $K=20$ in $300 \times 300 \mathrm{m^2}$ area. To get cumulative probability density functions (CDFs) of user rates, the performance evaluations are conducted 200 times with random deployments of APs and users.
    
For the large-scale fading, we adopt the 3-slope model \cite{Ao01Mobile} and the Cost-Hata model in \cite{Ngo17Cell-Free}. Thus, we can express the large scale fading $\beta_{mk}$ as follows:
\ifCLASSOPTIONonecolumn
  \begin{equation}
    [\beta_{mk}]_{\textrm{dB}} = - [ L_{\text{ch}}]_{\textrm{dB}}\nonumber\\ +
    \begin{cases}
      -35 \log_{10} (d_{mk}),  & \text{if} ~ d_{mk} > d_1, \\
      -15 \log_{10} (d_{1}) -20 \log_{10} (d_{mk}),  & \text{if} ~ d_0 < d_{mk} \leq d_1, \\
      -15 \log_{10} (d_{1}) -20 \log_{10} (d_{0}),  & \text{if} ~ d_{mk} \leq d_0, \nonumber
    \end{cases}
  \end{equation}
\else
  \begin{align}
    & [\beta_{mk}]_{\textrm{dB}} = - [ L_{\text{ch}}]_{\textrm{dB}}\nonumber\\
    & +
    \begin{cases}
      -35 \log_{10} (d_{mk}),  & \text{if} ~ d_{mk} > d_1, \\
      -15 \log_{10} (d_{1}) -20 \log_{10} (d_{mk}),  & \text{if} ~ d_0 < d_{mk} \leq d_1, \\
      -15 \log_{10} (d_{1}) -20 \log_{10} (d_{0}),  & \text{if} ~ d_{mk} \leq d_0, \nonumber
    \end{cases}
  \end{align}
\fi
where $d_{mk}$ represents the distance between the $m$th AP and the $k$th user in meter unit, and both $d_0$ and $d_1$ are parameters of the 3-slope path-loss model and set to 10m and 50m, respectively. In addition, $L_{\text{ch}}$ indicates the Cost-Hata radio propagation parameter that is given by
  \begin{align}
    [ L_{\text{ch}}]_{\textrm{dB}} = & ~ 46.3 + 33.9 \log_{10} (f) - 13.82 \log_{10} (h_{\textrm{AP}}) \nonumber \\
    & -(1.1 \log_{10} (f) - 0.7) h_{\textrm{user}} + 1.56 \log_{10} (f) - 0.8, \nonumber
  \end{align}
where $f$ [MHz] is the frequency of transmission, $h_{\textrm{AP}} [m]$ indicates the AP antenna effective height, and $h_{\textrm{user}} [m]$ represents the user antenna effective height. In the performance evaluations, we assume that $f=1.9 \textrm{GHz}$ with $20 \textrm{MHz}$ bandwidth $(B=20\textrm{MHz})$, $h_{\textrm{AP}}= 10 \textrm{m}$, and $h_{\textrm{User}}= 1.65 \textrm{m}$ \cite{Ngo17Cell-Free}. It is also assumed that $\tau_{\text{c}} = 200$ samples that consist of $\tau_{\text{p}} = 20$, $\tau_{\text{d}} = 90$, and $\tau_{\text{u}} = 90$ samples. The maximum transmission power per user is set to $\rho_{\text{u}} = \rho_{\text{t}} = 100 \textrm{mW}$ and the noise power is assumed to be $\sigma_w^2 = -92 \textrm{dBm}$.
     
  \subsection{Power Control}
   
  \begin{figure}
    \centering
    \includegraphics[width=\figwidth]{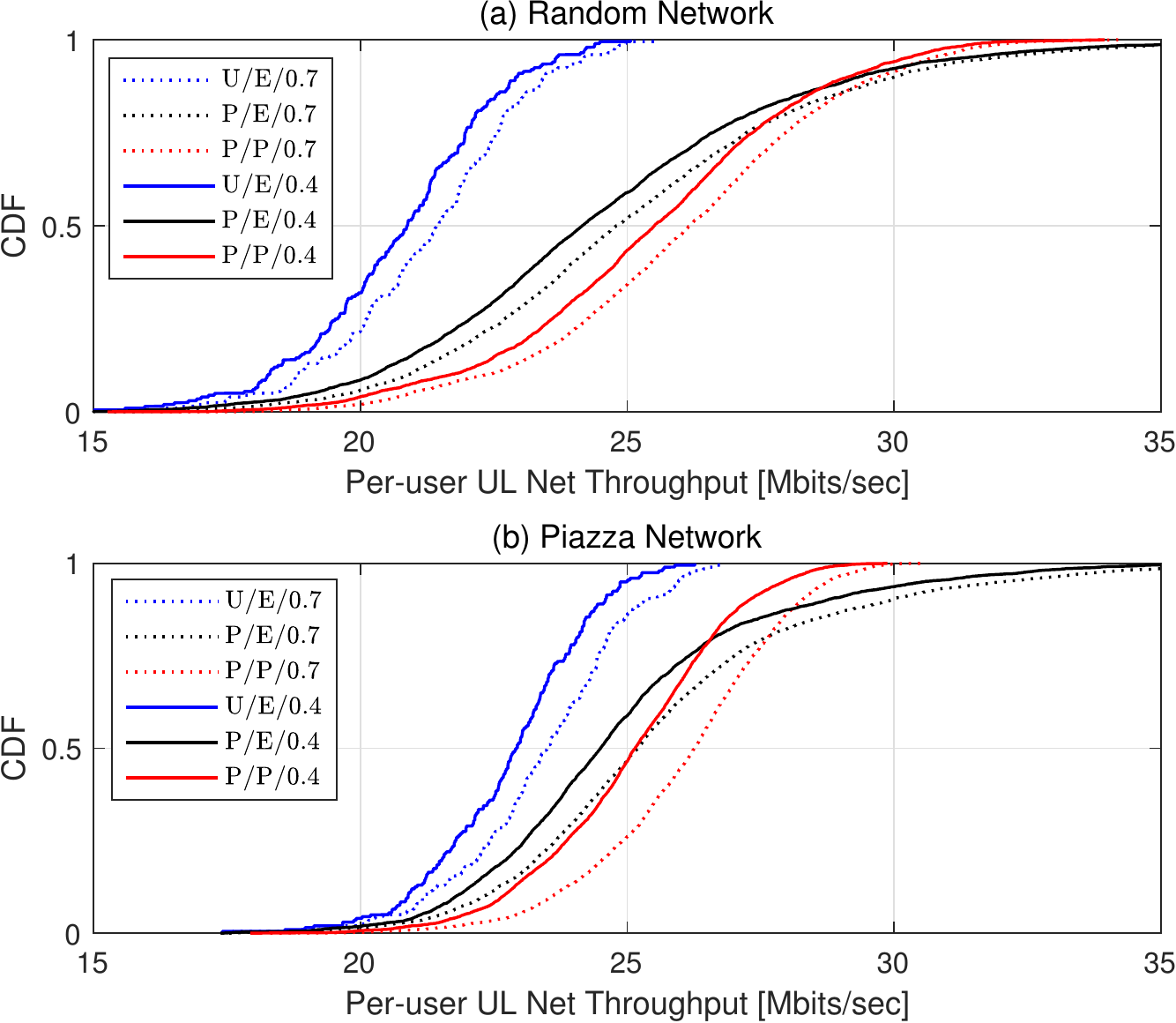}
    \caption{CDFs of throughput for $N=1$, $\xi_c = 0.4$ and $\xi_c = 0.7$ with power control and LLSF AP scheduling algorithm in random and piazza network.}
    \label{Fig:CDFAllocatedPower_Both}
  \end{figure}
    
In this section, we evaluate performances with the proposed and existing \cite{Ngo17Cell-Free, Guenach21Joint} power control algorithms while the LLSF algorithm \cite{Guenach21Joint} is adopted for the AP scheduling. Thus, we focus on the improvement due to the proposed power control algorithm. The performance evaluations are carried out in terms of per-user UL net throughput, simply called throughput hereafter, \cite{Ngo17Cell-Free} is defined as
  \[
    \bar{R}^{i}_k = B \left( 1 - \frac{\tau_{\text{p}}+\tau_{\text{d}}}{\tau_{\text{c}}}  \right) R^{i}_{k}, \quad  i\in \{ \text{Prop}, \text{UatF}\}. 
  \]
The CDFs of throughput are depicted in Fig. \ref{Fig:CDFAllocatedPower_Both} where the symbols $x$, $y$ $z$ in the legend $x/y/z$ tell the bounding technique, power control algorithm, and fronthaul constraint value, i.e., $\xi_c$. That is, P and U for $x$ indicate the proposed and UatF bounds, respectively, P and E for $y$ tell the proposed and existing algorithms, and $z$ represents the fronthaul constraint value. The curves in blue are the CDFs of throughput evaluated with the UatF bound. Meanwhile, the ones in red and black are obtained with the proposed bound. The comparisons between the proposed and the UatF bounds, i.e., P/E/$z$ vs. U/E/$z$ for $z = $ 0.7 and 0.4 clearly show that the UatF bound significantly underestimates the user rates. Now, we compare the throughput performances with the proposed and existing power control algorithm, i.e., P/P/$z$ vs. P/E/$z$ for $z = $ 0.7 and 0.4. The comparison demonstrates that the proposed algorithm considerably improves the $90\%$ and $95\%$-likely throughput in both networks. That is, the proposed power control algorithm improves the max-min fairness as compared to the existing one regardless of the network type. In particular, the comparison in Fig. \ref{Fig:CDFAllocatedPower_Both}(a) shows that the proposed power control algorithm with the proposed bound, i.e., P/P/$z$ improves the $95\%$-likely throughput about 21\% (20\%, resp.) with $z = 0.7$ ($z = 0.4$, resp.) as compared to U/E/$z$ in the random network. Meanwhile, for the piazza network, the comparison in Fig. \ref{Fig:CDFAllocatedPower_Both}(b) shows that  about 12\% (9\%, resp.) improvement in terms of $95\%$-likely throughput with $z = 0.7$ ($z = 0.4$, resp.) when P/P/$z$ is compared to U/E/$z$. It is also noticed that the performance improvements of the random network are larger than the ones in the piazza network. The reduced performance improvement of the piazza network tells that the channel hardening is not pronounced to more users in the random network since the variations of the distance between APs and a user are larger when APs are randomly deployed. 

  \subsection{Joint Power Control and AP Scheduling}

  \begin{figure}
    \centering
    \includegraphics[width=\figwidth]{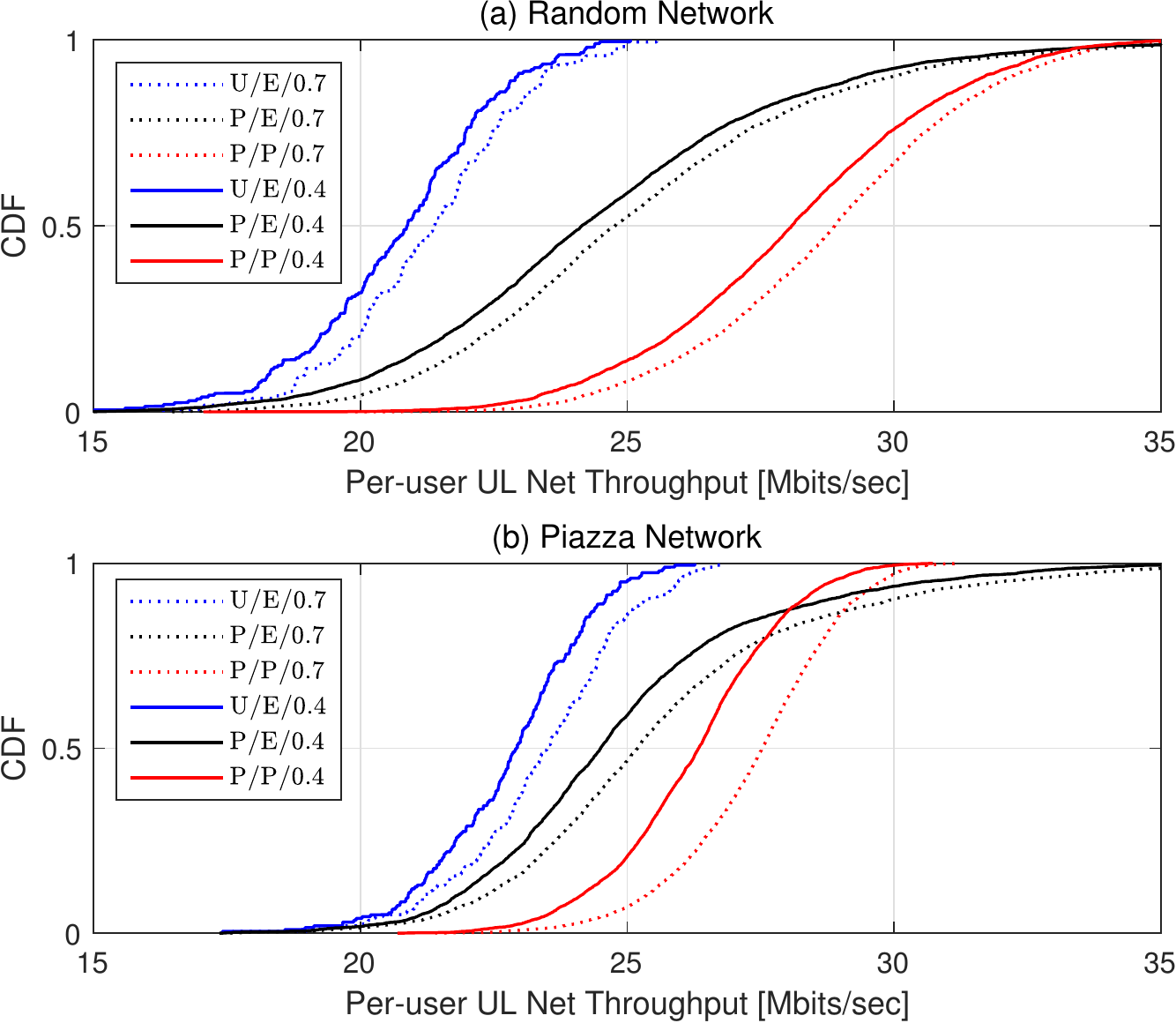}
    \caption{CDFs of throughput for $N=1$, $\xi_c = 0.4$ and $\xi_c = 0.7$ with joint power control and AP scheduling algorithm in random and piazza network.}
    \label{Fig:CDF_Comparision_UatF_Prop_JointOpt}
  \end{figure}
In this section, we evaluate the CDFs of throughput of CF-mMIMO network which is optimized with the proposed joint power control and AP scheduling algorithm. In addition, we also conduct the system optimization with the existing joint power control and AP scheduling algorithm in \cite{Guenach21Joint} and evaluate the CDFs of throughput for the designed systems. In particular, we take the IHB-based AP scheduling in \cite{Guenach21Joint} for the existing joint optimization. The performance evaluations are carried out for two fronthaul constraint values, i.e., $\xi_c = 0.7$ and 0.4, in Fig. \ref{Fig:CDF_Comparision_UatF_Prop_JointOpt} where the convention of legend follows the one in Fig. \ref{Fig:CDFAllocatedPower_Both}.

The results in Fig. \ref{Fig:CDF_Comparision_UatF_Prop_JointOpt} clearly show that the throughput performance is significantly underestimated when the system is designed with the existing algorithm and evaluated with the UatF bound. The systems designed with the proposed and existing algorithms are also compared in Fig. \ref{Fig:CDF_Comparision_UatF_Prop_JointOpt} based on the proposed bound. It is witnessed in the comparison that that the design with the proposed algorithm noticeably improves the fairness as compared to the one with the existing algorithm. In particular, from the comparison between P/P/$z$ and P/E/$z$ for the random network, it is observed in Fig. \ref{Fig:CDF_Comparision_UatF_Prop_JointOpt}(a) that the proposed algorithm, i.e., P/P/$z$, improves the 95\%-likely throughput about 23\% and 21\%, for $z = 0.4$ and 0.7, respectively as compared to P/E/$z$. For the piazza network, the fairness is also greatly improved with the proposed algorithm. For example, in the comparison between P/P/$z$ and P/E/$z$ shown in \ref{Fig:CDF_Comparision_UatF_Prop_JointOpt}(b), the proposed algorithm, i.e., P/P/$z$, has an improvement of 11\% and 15\%, for $z = 0.4$ and 0.7, respectively in terms of the 95\%-likely throughput. As observed in Fig. \ref{Fig:CDFAllocatedPower_Both}, the performance improvement of the random network is more noticeable than the piazza network.

  \begin{figure}[!t]
    \centering
    \includegraphics[width=\figwidth]{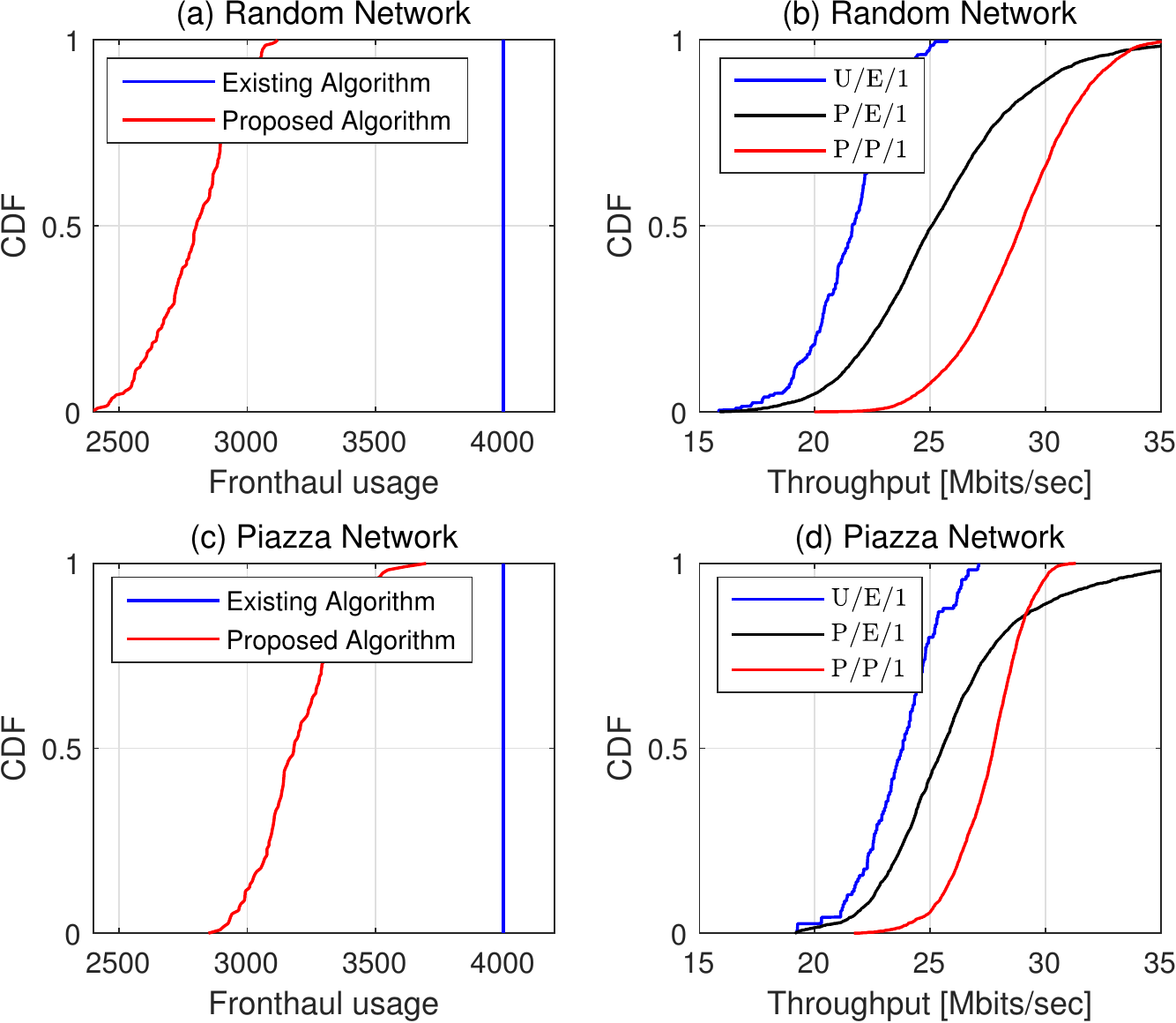}
    \caption{CDFs of the fronthaul usage, $\Xi$ and throughput for $\xi_c=1$ with joint power control and AP scheduling algorithm in random and piazza network.}
    \label{Fig:TotalNumUsedAP}
  \end{figure}

The study in \cite{Buzzi20User-Centric} discussed that in the uplink, the participation of all APs to the decoding of all users may results in performance degradation while requiring intolerably large bandwidth between APs and CPU. Then, it is a crucial question that how many APs should participate in serving a user to achieve the optimal max-min performance. Despite of the importance, the question has not been extensively studied yet. In \cite{Guenach21Joint}, the AP scheduling algorithm assumes each user is served by a fixed number of APs. Meanwhile, the proposed AP scheduling algorithm addresses the problem in such a way that the number of serving APs for a user is adaptively determined depending on the distances between APs and users, and transmit powers from users. In Fig. \ref{Fig:TotalNumUsedAP}, we measure the CDFs of the fronthaul usage, $\Xi$ in \eqref{eq:usage} for the random and piazza networks when the fronthaul constraint is set to unity, i.e., $\xi_c = 1$. In the case of the existing algorithm, all users are served by the same number of APs. In particular, there are 200 APs and 20 users in the networks. Thus, the fronthaul usage amounts to $\Xi = 4,000$ ($= 200 \times 20$) since we set the fronthaul constraint value to unity. Meanwhile, the proposed algorithm adaptively assigns APs to each user depending on the network situations and requires a much reduced fronthaul usage, which is clear benefit when the franthaul bandwidth is limited. In Figs. \ref{Fig:TotalNumUsedAP}(a) and (c), it is observed that the proposed algorithm utilizes only 70\% and 80\% of the available fronthaul resources, i.e., 4,000, for the random and piazza networks, respectively. While the proposed algorithm requires the reduced fronthaul usage, it is observed in Figs. \ref{Fig:TotalNumUsedAP}(b) and (d) that noticeable performance improvement is gained in terms of fairness for both the random and piazza networks, respectively. 

  \begin{figure}[!t]
    \centering
    \includegraphics[width=\figwidth]{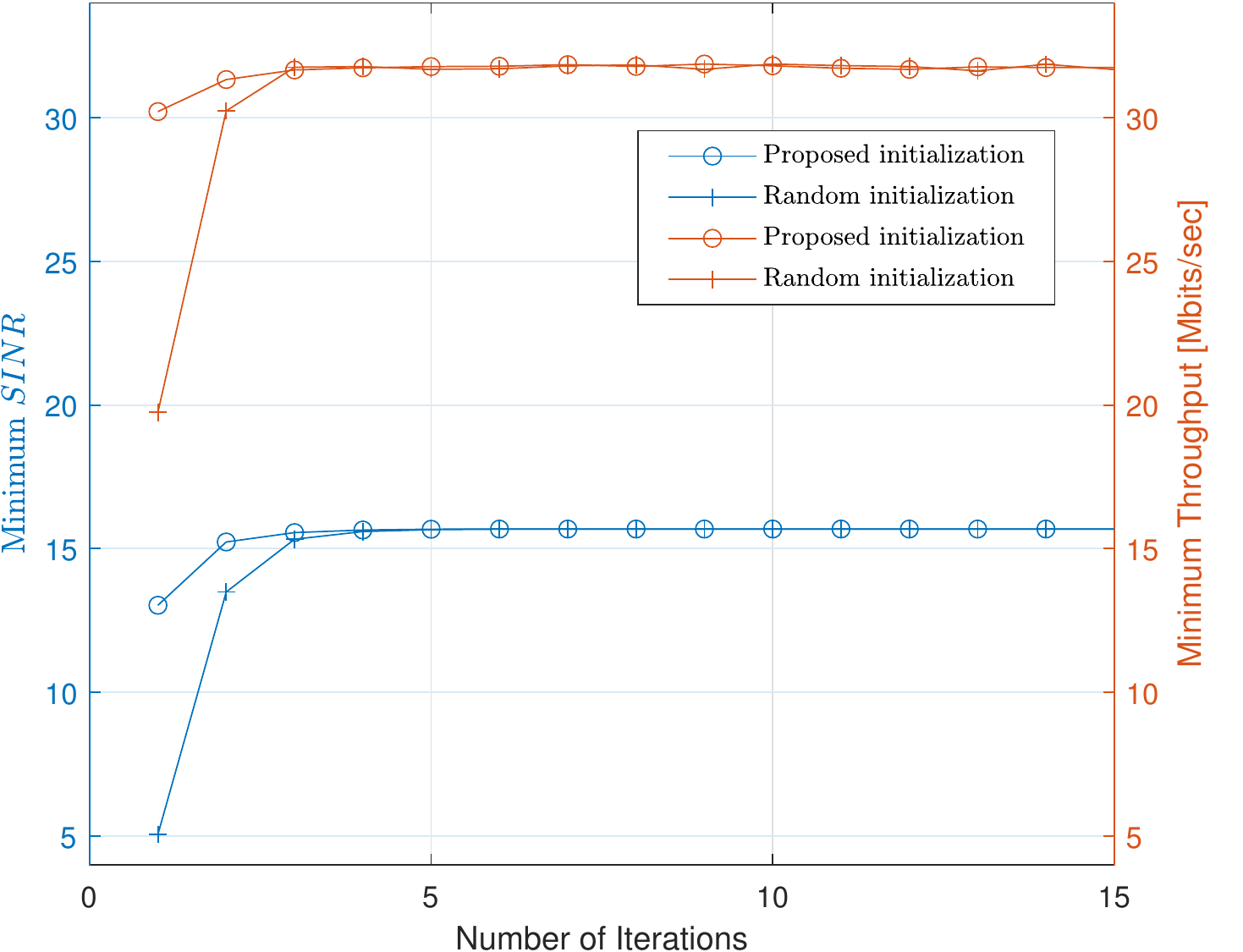}
    \caption{Convergence behavior of joint power control and AP scheduling with $M=200, N=1, K=20$ and $\xi_c=0.7$.}
    \label{Fig:ConvBehav}
  \end{figure}

Fig. \ref{Fig:ConvBehav} shows both the minimum SINR, i.e., $\min_k{\text{SINR}_k^{\text{Prop}}}$ and minimum throughput, i.e., $\min_k{\bar{R}^{\text{Prop}}_k}$ for $\xi_c = 0.7$ at each iteration of Algorithm \ref{Alg:JointAOFrame} when the initial AP connection coefficients are randomly established and taken from the LLSF algorithm. It is noticed that the proposed algorithm makes both the minimum SINR and the minimum throughput converge to fixed values. In addition, the proposed initialization, i.e., the initial AP connections with the LLSF algorithm, accelerates the convergence.

\section{Conclusion} \label{Sec:Conclusion}

In this work, we propose a new lower bound on the achievable rate for UL CF-mMIMO system without assuming the channel hardening. In addition,  we develop a joint power control and AP scheduling algorithm based on the proposed bound. Comparisons between performance estimates with the proposed and UatF bounds reveal that the existing UaF bound  often seriously underestimates the system performance when the channel hardening is not prominent. It is discussed that the inaccurate estimate leads to inefficient resource allocations, and thus system designs based on the UatF bound provide suboptimal design results. On the contrary, thanks to a tighter bound derived in the paper, we demonstrate that the proposed joint power control and AP scheduling algorithm provides noticeable improvement of fairness at much reduced bandwidth between APs and CPU.

\appendices

\section{Proof of Proposition \ref{Prop:Bound}} \label{Sec:ProofofProp1}
Here, we derive the lower bound in \eqref{Eq:ProposedBound}. By omitting the scaling factor and applying the chain rule, we have the mutual information in \eqref{Eq:ExactRate} as
\ifCLASSOPTIONonecolumn
  \begin{align}
    I&(q_k [1:\tau_{\text{u}}] ; r_k[1:\tau_{\text{u}}]) = I(\{ \mathbf{g}_{k,k'}, \hat{\mathbf{g}}_{k,k'} ; \forall k'\} ; r_k[1:\tau_{\text{u}}]) \nonumber \\
    & \qquad + I(q_k [1:\tau_{\text{u}}] ; r_k[1:\tau_{\text{u}}] | \{ \mathbf{g}_{k,k'}, \hat{\mathbf{g}}_{k,k'} ; \forall k'\} ) - I( \{ \mathbf{g}_{k,k'}, \hat{\mathbf{g}}_{k,k'} ; \forall k'\} ; r_k[1:\tau_{\text{u}}] | q_k [1:\tau_{\text{u}}] ). \label{Eq: Chainrule}
  \end{align}
\else
  \begin{align}
    I&(q_k [1:\tau_{\text{u}}] ; r_k[1:\tau_{\text{u}}]) = I(\{ \mathbf{g}_{k,k'}, \hat{\mathbf{g}}_{k,k'} ; \forall k'\} ; r_k[1:\tau_{\text{u}}]) \nonumber \\
    & \qquad + I(q_k [1:\tau_{\text{u}}] ; r_k[1:\tau_{\text{u}}] | \{ \mathbf{g}_{k,k'}, \hat{\mathbf{g}}_{k,k'} ; \forall k'\} ) \nonumber \\
    & \qquad \quad - I( \{ \mathbf{g}_{k,k'}, \hat{\mathbf{g}}_{k,k'} ; \forall k'\} ; r_k[1:\tau_{\text{u}}] | q_k [1:\tau_{\text{u}}] ). \label{Eq: Chainrule}
  \end{align}
\fi

Since the first term in the right hand side of \eqref{Eq: Chainrule} is positive, by removing the first term, we can a lower bound as follows:
\ifCLASSOPTIONonecolumn
  \begin{align}
    I(q_k & [1:\tau_{\text{u}}] ; r_k[1:\tau_{\text{u}}]) \nonumber \\
    \geq & I(q_k [1:\tau_{\text{u}}] ; r_k[1:\tau_{\text{u}}] | \{ \mathbf{g}_{k,k'}, \hat{\mathbf{g}}_{k,k'} ; \forall k'\} ) - I( \{ \mathbf{g}_{k,k'}, \hat{\mathbf{g}}_{k,k'} ; \forall k'\} ; r_k[1:\tau_{\text{u}}] | q_k [1:\tau_{\text{u}}] ). \label{Ineq:MutualLower}
  \end{align}
\else
  \begin{align}
    I(q_k & [1:\tau_{\text{u}}] ; r_k[1:\tau_{\text{u}}]) \nonumber \\
    \geq & I(q_k [1:\tau_{\text{u}}] ; r_k[1:\tau_{\text{u}}] | \{ \mathbf{g}_{k,k'}, \hat{\mathbf{g}}_{k,k'} ; \forall k'\} ) \nonumber \\
    & - I( \{ \mathbf{g}_{k,k'}, \hat{\mathbf{g}}_{k,k'} ; \forall k'\} ; r_k[1:\tau_{\text{u}}] | q_k [1:\tau_{\text{u}}] ). \label{Ineq:MutualLower}
  \end{align}
\fi
When the instantaneous channel state information, i.e., $\mathbf{g}_{k,k'}$ and $\hat{\mathbf{g}}_{k,k'}$, is given, and the information signal is Gaussian, the mutual information between $q_k [1:\tau_{\text{u}}]$ and $r_k[1:\tau_{\text{u}}]$ can be expressed as
  \begin{align}
    & I(q_k [1:\tau_{\text{u}}] ; r_k[1:\tau_{\text{u}}] | \{ \mathbf{g}_{k,k'}, \hat{\mathbf{g}}_{k,k'} ; \forall k'\} ) \nonumber \\
    & = \tau_{\text{u}}\mathbb{E} \left[ \log_2 \left( 1 + \frac{|f_{k,k}|^2}{ \sum_{k' \neq k} |f_{k,k'}|^2 + \sum_{m=1}^{M} | c_{mk} \hat{\mathbf{g}}_{mk}^{H}|^2  \sigma_w^2} \right) \right]. \nonumber
  \end{align}
Now, by finding an upper bound on the second term in the right hand side of \eqref{Ineq:MutualLower}, we can have a lower bound on the achievable rate in \eqref{Eq:ExactRate}. By the definition of mutual information, we have the following eqaulity:
\ifCLASSOPTIONonecolumn
  \begin{align}
    I( & \{ \mathbf{g}_{k,k'}, \hat{\mathbf{g}}_{k,k'} ; \forall k'\} ; r_k[1:\tau_{\text{u}}] | q_k [1:\tau_{\text{u}}] )  \nonumber\\
    & = h(r_k[1:\tau_{\text{u}}] | q_k [1:\tau_{\text{u}}]) - h(r_k[1:\tau_{\text{u}}] | q_k [1:\tau_{\text{u}}], \{ \mathbf{g}_{k,k'}, \hat{\mathbf{g}}_{k,k'} ; \forall k'\}), \label{eq:mutualinfo}
  \end{align}
\else
  \begin{align}
    I( & \{ \mathbf{g}_{k,k'}, \hat{\mathbf{g}}_{k,k'} ; \forall k'\} ; r_k[1:\tau_{\text{u}}] | q_k [1:\tau_{\text{u}}] )  \nonumber\\
    & = h(r_k[1:\tau_{\text{u}}] | q_k [1:\tau_{\text{u}}])  \nonumber\\
    & \quad\quad - h(r_k[1:\tau_{\text{u}}] | q_k [1:\tau_{\text{u}}], \{ \mathbf{g}_{k,k'}, \hat{\mathbf{g}}_{k,k'} ; \forall k'\}), \label{eq:mutualinfo}
  \end{align}
\fi
where the first term in the right hand side  is maximized when the elements in the random vector, $r_k[1:\tau_{\text{u}}]$ are independent Gaussian random variables given $q_k [1:\tau_{\text{u}}]$. Thus, we have the following upper bound for the first term:
\ifCLASSOPTIONonecolumn
  \begin{align}
    h( & r_k[1:\tau_{\text{u}}] | q_k [1:\tau_{\text{u}}]) \nonumber \\
    & \leq \tau_{\text{u}} \log_2 \left( \sum_{k'\neq k} \mathbb{E} \left[ \left|f_{k,k'} \right|^2 \right] + \mathbb{E} \left[ \sum_{m=1}^{M} \left| c_{mk}  \hat{\mathbf{g}}_{mk}^{H} \right|^2 \right]  \sigma_w^2 \right) \nonumber\\
    & + \log_2 \left( 1 + \frac{ \tau_{u}  \mathbb{V} [f_{k,k}] }{\sum_{k'\neq k} \mathbb{E} [ |f_{k,k'}|^2 ] + \mathbb{E} [ \sum_{m=1}^{M} |  c_{mk}  \hat{\mathbf{g}}_{mk}^{H}|^2 ]  \sigma_w^2} \right) + \tau_{\text{u}} \log_2 (e \pi) . \nonumber
  \end{align}
\else
  \begin{align}
    h( & r_k[1:\tau_{\text{u}}] | q_k [1:\tau_{\text{u}}]) \nonumber \\
    & \leq \tau_{\text{u}} \log_2 \left( \sum_{k'\neq k} \mathbb{E} \left[ \left|f_{k,k'} \right|^2 \right] + \mathbb{E} \left[ \sum_{m=1}^{M} \left| c_{mk}  \hat{\mathbf{g}}_{mk}^{H} \right|^2 \right]  \sigma_w^2 \right) \nonumber\\
    & + \log_2 \left( 1 + \frac{ \tau_{u}  \mathbb{V} [f_{k,k}] }{\sum_{k'\neq k} \mathbb{E} [ |f_{k,k'}|^2 ] + \mathbb{E} [ \sum_{m=1}^{M} |  c_{mk}  \hat{\mathbf{g}}_{mk}^{H}|^2 ]  \sigma_w^2} \right) \nonumber \\
    & \hspace{0.6\columnwidth} + \tau_{\text{u}} \log_2 (e \pi) . \nonumber
  \end{align}
\fi
Please refer to \cite{Caire18OnTheErgodic} about the specific mathematical manipulations. For the second term in the right hand side of \eqref{eq:mutualinfo}, the received signal, $r_k[1:\tau_{\text{u}}]$, is Gaussian given the data signal and effective channels, $q_k [1:\tau_{\text{u}}],$ and $\{\mathbf{g}_{k,k'}, \hat{\mathbf{g}}_{k,k'} ; \forall k'\}$. Thus, the second term can be readily given by 
\ifCLASSOPTIONonecolumn
  \begin{align*}
    h( & r_k[1:\tau_{\text{u}}] | q_k [1:\tau_{\text{u}}], \{ \mathbf{g}_{k,k'}, \hat{\mathbf{g}}_{k,k'} ; \forall k'\}) \nonumber\\
    & = \tau_{\text{u}}  \mathbb{E} \left[  \log_2 \left( \sum_{k'\neq k} \left|f_{k,k'} \right|^2  + \sum_{m=1}^{M} \left| c_{mk}  \hat{\mathbf{g}}_{mk}^{H} \right|^2  \sigma_w^2 \right) \nonumber \right] + \tau_{\text{u}} \log_2 (e \pi). 
  \end{align*}
\else
  \begin{align*}
    h( & r_k[1:\tau_{\text{u}}] | q_k [1:\tau_{\text{u}}], \{ \mathbf{g}_{k,k'}, \hat{\mathbf{g}}_{k,k'} ; \forall k'\}) \nonumber\\
    & = \tau_{\text{u}}  \mathbb{E} \left[  \log_2 \left( \sum_{k'\neq k} \left|f_{k,k'} \right|^2  + \sum_{m=1}^{M} \left| c_{mk}  \hat{\mathbf{g}}_{mk}^{H} \right|^2  \sigma_w^2 \right) \nonumber \right] \\
    & \hspace{.6\columnwidth} + \tau_{\text{u}} \log_2 (e \pi).
  \end{align*}
\fi
\qed

\section{Proof of Proposition \ref{Prop:ApproxUpp}} \label{Sec:ProofofProp2}
Let us first apply the Jensen's inequality to the term (a) of the proposed bound in \eqref{Eq:ProposedBound}. Then, we get an upper bound on the proposed bound in \eqref{Eq:Prop3_Jen}.
  \begin{figure*}[t!]
\ifCLASSOPTIONonecolumn
  \begin{align}
    R_k^{\text{Prop}} 
    & \leq  \log_2 \left(  1 + \frac{\mathbb{E} [|f_{k,k}|^2 ] }{\sum_{k'\neq k} \mathbb{E} [ |f_{k,k'}|^2 ] + \mathbb{E} [ \sum_{m=1}^{M}  | c_{mk}  \hat{\mathbf{g}}_{mk}^{H}|^2 ]  \sigma_w^2} \right) \nonumber\\
    & - \frac{1}{\tau_{u}} \log_2 \left( 1 + \frac{ \tau_{u}  \mathbb{V} [f_{k,k}] }{\sum_{k'\neq k} \mathbb{E} [ |f_{k,k'}|^2 ] + \mathbb{E} [ \sum_{m=1}^{M} | c_{mk}  \hat{\mathbf{g}}_{mk}^{H}|^2 ]  \sigma_w^2} \right). \label{Eq:Prop3_Jen}
  \end{align}
\else
  \begin{align}
    R_k^{\text{Prop}} 
    & \leq  \log_2 \left(  1 + \frac{\mathbb{E} [|f_{k,k}|^2 ] }{\sum_{k'\neq k} \mathbb{E} [ |f_{k,k'}|^2 ] + \mathbb{E} [ \sum_{m=1}^{M}  | c_{mk}  \hat{\mathbf{g}}_{mk}^{H}|^2 ]  \sigma_w^2} \right) \nonumber\\
    &\hspace{0.7\columnwidth} - \frac{1}{\tau_{u}} \log_2 \left( 1 + \frac{ \tau_{u}  \mathbb{V} [f_{k,k}] }{\sum_{k'\neq k} \mathbb{E} [ |f_{k,k'}|^2 ] + \mathbb{E} [ \sum_{m=1}^{M} | c_{mk}  \hat{\mathbf{g}}_{mk}^{H}|^2 ]  \sigma_w^2} \right). \label{Eq:Prop3_Jen}
  \end{align}
\fi
    \hrulefill
  \end{figure*}
Now, we approximate the effective channel gain, $f_{k,k}$, following the technique introduced in \cite{Interdonato21Enhanced} as
  \begin{equation}
    f_{k,k}  = \sum_{m=1}^{M} \sqrt{\rho_{\text{u}}  \eta_{k}}  c_{mk}  \mathbf{\hat{g}}_{mk}^H   \mathbf{g}_{mk} 
    \approx \sum_{m=1}^{M} \sqrt{\rho_{\text{u}}  \eta_{k}}  c_{mk}  \mathbf{g}_{mk}^H   \mathbf{g}_{mk}. \label{Eq:NoChErr}
  \end{equation}

This approximation is plausible when we have enough communication resources to allocate an orthogonal pilot sequence to each user or use an intelligent pilot allocation scheme to minimize the pilot contamination such as graph coloring based pilot assignment \cite{Liu20Graph}. The approximation of effective channel gain in \eqref{Eq:NoChErr} is the summation of multiple random variables that follow gamma distributions with different scale factors, i.e., $\beta_{mk}$ but the same shape parameter, i.e., $N$. By using the Welch Satterthwaite approximation \cite{Zheng21Wireless, Abusabah21Approximate}, we can approximate the summation of independent non-identically distributed Gamma random variables as a single Gamma random variable whose first and second moments are the same as those of the sum of random variables, i.e. the right-hand side of \eqref{Eq:NoChErr}. Then, the approximation allows us to have the CDF of the effective channel gain as

\[
    \textrm{Pr} [f_{k,k} \leq f] = \frac{1}{\Gamma(v_k)} \gamma \left( v_k,\frac{f}{w_k} \right),  
\]
where $\Gamma ( \cdot )$ is the Gamma function, $\gamma ( \cdot )$ is the lower incomplete gamma function, and
\[
    v_k = \frac{(\sum_{m=1}^{M} c_{mk}  N  \beta_{mk})^2}{\sum_{m=1}^{M} c_{mk}^2  N  \beta_{mk}^2}, ~
    w_k = \sqrt{\rho_{\text{u}}  \eta_k} \frac{\sum_{m=1}^{M} c_{mk}^2  \beta_{mk}^2}{\sum_{m=1}^{M} c_{mk}  \beta_{mk}},
\]
 are the shape and scale parameters of the Gamma distribution, respectively. By using the property of Gamma distribution, we can get a relation between the variance and the second moment of Gamma distribution as follows
  \begin{equation}
    \mathbb{V} [f_{k,k}] = \frac{\mathbb{E} [|f_{k,k}|^2]}{1+v_k}, \label{Eq:Approx_Var}
  \end{equation}
which is lower bounded as 
  \begin{equation}
    \mathbb{V} [f_{k,k}] \ge  \frac{\mathbb{E} [|f_{k,k}|^2]}{1+MN}, \label{Eq:ApproxUpper_Var}
  \end{equation}
since $v_k$ in \eqref{Eq:Approx_Var} is upper bounded as
  \begin{equation}
    v_k = \frac{(\sum_{m=1}^{M} c_{mk}  N  \beta_{mk})^2}{\sum_{m=1}^{M} c_{mk}^2  N  \beta_{mk}^2} \leq NM, \label{Eq:Upper_v_k}
  \end{equation}
due to the Cauchy-Schwarz inequality. Finally, by replacing $\mathbb{V} [f_{k,k}]$ in \eqref{Eq:Prop3_Jen} with the lower bound in \eqref{Eq:ApproxUpper_Var}, the bound in \eqref{InEq:ApproxUpper} follows. \qed
 

\section{Proof of Proposition \ref{Prop:QuasiConcave}} \label{Sec:ProofofProp3}
To prove the quasi-concavity, let us first define the upper-level set of the objective function in \eqref{Opt:PowAll} as follows:
  \begin{align}
    U(t) & = \{ \eta ; \min_{k} \text{SINR}_k^{\text{Prop} }  \geq t \} \nonumber \\
    & = \{ \eta ; \text{SINR}_k^{\text{Prop} }  \geq t , \forall k \} \nonumber \\
    & = \left\{ \eta ;  \eta^{T} \mathbf{z}_{k} \geq t \sum_{m=1}^{M} c_{mk}  \gamma_{mk}  N  \sigma_w^2, \forall k \right\} \label{Eq:UpperSetProp3}
  \end{align}
where $\eta  = [\eta_{1},\eta_{2}, \ldots, \eta_{K}]^{T} $,  $\mathbf{z}_{k} = [z_{k,1}, z_{k,2}, \ldots, z_{k,K}]^T$, and
\[
z_{k,i} =        
  \begin{cases}
    \rho_{\text{u}}  \left[\sum_{m=1}^{M} c_{mk}  N  \gamma_{mk}  \beta_{mk} + \left(\sum_{m=1}^{M} c_{mk}  N  \gamma_{mk} \right)^2 \right],  \\ \qquad     \qquad \qquad \qquad \qquad \qquad \qquad \qquad \qquad \text{if} ~ i = k, \\
     -t \rho_{\text{u}}  \left[\sum_{m=1}^{M} c_{mk}  N  \gamma_{mk}  \beta_{mi} \right], \quad \text{if} ~ i \neq k.
  \end{cases}
\]
For all $t$, the elements of the upper-level set in \eqref{Eq:UpperSetProp3} are in the half-space divided by the following hyperplane:
\[
  \eta^{T} \mathbf{z}_{k} = t \sum_{m=1}^{M} c_{mk}  \gamma_{mk}  N  \sigma_w^2.
\]

In such a case, the upper-level set of the objective function in \eqref{Opt:PowAll} is a convex set, and thus the objective function is a quasi-concave function of $\eta$ \cite[Section 2.2.1]{Boyd04Convex}. \qed


\section{Proof of Proposition \ref{Prop:QuasiConcave_APsch}} \label{Sec:ProofofProp4}
Similar to the Appendix \ref{Sec:ProofofProp3}, let us first define the upper-level of the objective function in \eqref{Opt:APSch_RelaxMM} set as follows:
  \begin{align}
    U(t) & = \{ [ \mathbf{c}_{1}, \mathbf{c}_{2}, \ldots, \mathbf{c}_{K} ] ; \min_{k} g( \mathbf{c}_{k} ; \mathbf{c}_{k}^{[n]} )  \geq t \} \nonumber \\
    & = \{ [ \mathbf{c}_{1}, \mathbf{c}_{2}, \ldots, \mathbf{c}_{K} ] ; g( \mathbf{c}_{k} ; \mathbf{c}_{k}^{[n]} )  \geq t , \forall k \} \nonumber  \\
    & = \left\{ [ \mathbf{c}_{1}, \mathbf{c}_{2}, \ldots, \mathbf{c}_{K} ] ;   \mathbf{c}_{k}^T \mathbf{z}_k^{[n]}  \geq t y_k^{[n]} - w_k^{[n]} , \forall k \right\}, \label{Eq:UpperSetProp4}
  \end{align}
where
  \begin{align*}
    \mathbf{z}_k^{[n]} = \eta_{k} ( \bm{\gamma}_{k} \otimes {\boldsymbol{\beta}}_{k})^{T} + 2 N  \eta_{k} \mathbf{c}_{k}^{[n]}, \\
    y_k=  \sum_{k' \neq k} \eta_{k'}  ( \bm{\gamma}_{k} \otimes \bm{\beta_{k'}})^T  + \frac{\sigma_w^2}{\rho_{\text{u}}} \bm{\gamma_{k}}^T, \\
    w_k^{[n]} = N  \eta_{k}  \left( \left( \bm{\gamma}_{k}^{T}    \mathbf{c}_{k}^{[n]} \right)^2 - 2 (\mathbf{c}_{k}^{[n]})^2  \right).
  \end{align*}
Since the set in \eqref{Eq:UpperSetProp4} is a half-space, the set is convex, which means that the objective function in \eqref{Opt:APSch_RelaxMM} is quasi-concave. In addition, the sets constructed with the constraints in \eqref{Opt:APSch_RelaxMM}, i.e., \eqref{Eq:FHconst}, \eqref{Eq:AddrelaxConst}, and \eqref{Eq:RelaxConst} are convex. Therefore, the problem in \eqref{Opt:APSch_RelaxMM} is a quasi-concave optimization problem. \qed


\ifCLASSOPTIONcaptionsoff
  \newpage
\fi

\bibliographystyle{IEEEtran}

\end{document}